\documentclass[12pt]{article}
\pdfoutput=1
\usepackage{putex}
\usepackage{graphicx}
\usepackage{epstopdf}
\usepackage{enumerate}
\usepackage[nosort]{cite}
\usepackage{tensor}
\usepackage{slashed}
\usepackage{feynmf}
\usepackage{hyperref}
\usepackage{float}
\usepackage[bottom]{footmisc}
\usepackage[utf8]{inputenc}
\usepackage{amsmath}
\usepackage[usenames,dvipsnames,svgnames,table]{xcolor}
\usepackage{tikz}
\usetikzlibrary{positioning}
\usetikzlibrary{decorations.pathmorphing}
\usepackage{tikz-feynman}
\usepackage{tikz-feynhand}
\usepackage[bf,hang,footnotesize]{caption}
\usepackage{bbold}

\numberwithin{equation}{section}
\hypersetup{
	pdfstartview={FitH},    
	pdftitle={Carrollian Partition Function for Bulk Yang-Mills Theory},    
	pdfauthor={Kraus, Myers},     
	colorlinks=true,       
	linkcolor=blue,          
	citecolor=blue!59!black! ,        
	filecolor=magenta,      
	urlcolor=blue!75!black,           
	linktoc=all
}

\makeatletter
\g@addto@macro\bfseries{\boldmath}
\makeatother

\numberwithin{equation}{section}

\newcommand{\eq}[2]{\begin{align}\label{#1}#2\end{align}}

\newcommand {\be} {\begin {equation}}
\newcommand {\ee} {\end {equation}}

\newcommand{\p}{\partial}

\def\tr{\mathop{\rm tr}}
\newcommand\rt{{\rightarrow}}
\def\eps{\epsilon}

\newcommand{\rf}[1]{(\ref{#1})}
\newcommand{\rff}[1]{\ref{#1}}

\newcommand{\zb}{\overline{z}}

\newcommand{\veps}{\varepsilon}

\newcommand{\arrow}{\rightarrow}
\newcommand{\hatline}[1]{\hat{\overline{#1}}}
\newcommand{\A}{\mathcal{A}}

\definecolor{greenC}{rgb}{0.0, 0.38, 0.18}

\newcommand{\Ab}{\overline{A}}

\newcommand{\xh}{\hat{x}}

\newcommand{\struct}[3]{
f^{#1}_{\ \,\,#2#3}
}

\newcommand{\dr}{d}

\newcommand{\scrI}{\mathcal I}
\newcommand{\pv}{\vec{p}}

\newcommand{\Ic}{{\cal I}}

\newcommand{\nabh}{\hat{\nabla}}

\newcommand{\D}{\mathcal{D}}

\newcommand{\qv}{\vec{q}}
\newcommand{\qh}{\hat{q}}
\newcommand{\yh}{\hat{y}}
\newcommand{\subsuper}[2]{{^{#2}_{#1}}}
\newcommand{\vepsh}{\hat{\varepsilon}}

\begin{document}

\institution{UCLA}{ \quad\quad\quad\quad\quad\quad\quad\ ~ \, $^{1}$Mani L. Bhaumik Institute for Theoretical Physics
		\cr Department of Physics \& Astronomy,\,University of California,\,Los Angeles,\,CA\,90095,\,USA}

\title{Carrollian Partition Function for \\ Bulk Yang-Mills Theory  }

\authors{Per Kraus$^{1}$, Richard M. Myers$^{1}$}
	
\abstract{ 
The path integral over massless quantum fields in Minkowski space with scattering boundary conditions defines a Carrollian   partition function on the null boundary.  We develop this framework for  non-Abelian gauge theory, both from a general perspective and through explicit examples that highlight  subtle aspects of soft modes and asymptotic symmetries.  These  include falloff conditions, Goldstone modes and their antipodal matching, and factors of two  associated with conditionally convergent integrals arising in the derivation of soft theorems.  We employ path integral (rather than canonical) methods throughout.
 }
	
	\date{}
	
	\maketitle
	\setcounter{tocdepth}{2}
	\begingroup
	\hypersetup{linkcolor=black}
	\tableofcontents
	\endgroup
	

\section{Introduction}

The development of the holographic principle has been a  boon to our  understanding of quantum gravity. The AdS/CFT correspondence is the best understood realization of this principle, and many tools have been developed in that context.  Much effort has naturally been devoted to trying to replicate this success in asymptotically flat spacetime.   

Holography is, at its core, a statement about gravity. However, one of the lessons of AdS/CFT is that much can be learned about the holographic dictionary by considering non-gravitational theories in the bulk via  a bottom-up approach. 
The work presented here is in that spirit, continuing the line of research initiated in \cite{Kraus:2024gso, Kim:2023qbl},  which aims to provide a path integral derivation of the dictionary for a (presently hypothetical) holographic duality for flat spacetime. Specifically, we describe the path integral dictionary for the case of bulk non-Abelian gauge fields, both as a warmup for the more complicated case of bulk gravity and as a useful venue  for elucidating and streamlining various points in prior work.

The two major approaches to flat space holography lie under the rubrics of Carrollian\footnote{Various aspects of Carrollian symmetries and field theories have been studied in \cite{Duval:2014uva,Duval:2014uoa,Hartong:2015xda,Hartong:2015usd,Bagchi:2019clu,Bagchi:2016bcd,Ciambelli:2019lap,Ciambelli:2018wre,deBoer:2023fnj,Nguyen:2023vfz,Cotler:2024xhb}. See e.g. \cite{Donnay:2022aba,Bagchi:2022emh,Donnay:2022wvx,Bagchi:2023fbj,Mason:2023mti,Alday:2024yyj} for work on the Carrollian perspective on flat space holography.} and celestial\footnote{In celestial holography, the dual theory instead lives on the celestial 2-sphere. The literature on celestial holography is now extensive, and we refer the reader to the selection of reviews \cite{Strominger:2017zoo,Pasterski:2021rjz,Raclariu:2021zjz,Pasterski:2021raf, McLoughlin:2022ljp,Donnay:2023mrd}. The relation between celestial and Carrollian correlators has been worked out in many places, e.g.  \cite{Donnay:2022aba,Bagchi:2022emh,Donnay:2022wvx,Bagchi:2023cen}.  Celestial holography motivated treatments of Yang-Mills theory in the covariant phase space formalism include  \cite{Strominger:2013lka,He:2020ifr,Campiglia:2021oqz,He:2023bvv}.} holography, typically in a bottom-up framework.\footnote{Alternative holographic approaches from a top down perspective include \cite{Costello:2022wso,Costello:2022jpg,Costello:2023hmi}.} In this work we will focus on the Carrollian approach, in which the S-matrix elements of massless bulk fields are described by the correlators of a putative Carrollian CFT supported on null infinity.  Many basic issues remain open, and a sampling of questions --- mostly involving the handling of asymptotic symmetries ---  that motivated the present work include:
\begin{itemize}
    \item Large gauge symmetry appears to require the presence of certain antipodal matching conditions on the large gauge Goldstone.\footnote{To avoid confusion, here we refer to antipodal matching of the Goldstone (proposed in \cite{He:2014cra}), not of the so-called soft creation operator whose matching is discussed in \cite{Strominger:2017zoo}.} Are these conditions gauge dependent? Do they survive quantum corrections? Does the large gauge symmetry itself survive quantization, or do central extensions or other deformations to the charge algebra develop at loop order?

    \item Many treatments of large gauge symmetry involve the introduction of a so-called soft sector composed of the Goldstone of spontaneously broken large gauge symmetry and its conjugate. How, precisely, does this sector relate to the ``hard" phase space of finite-energy asymptotic particles in standard treatments of scattering?

    \item Derivations of the asymptotic charge algebra often proceed classically and finiteness of the charges requires assumptions about the behavior of the fields near the bounds of null infinity, i.e. near spatial and timelike infinities. Are these assumptions valid, even perturbatively,  in the context of quantum scattering?

    \item Obtaining the correct coefficient in the soft theorem from the Ward identity for large gauge transformations involves a subtle factor of $2$, usually introduced in the canonical formalism via a rule for computing Dirac brackets \cite{He:2014laa,He:2014cra}, but deserves clarification.   In the path integral formalism employed here the derivation of the soft theorem encounters a subtlety involving the appearance  of integrals that are only conditionally convergent.  These two subtleties are related, and we'll see that the natural (Poincar\'{e}  invariant) definition of such integrals allows for a transparent treatment of these factors of $2$.    
\end{itemize}

Our goal in this work is to develop tools, inspired by those familiar from AdS/CFT, for addressing questions like these. Below we will describe in more detail our precise findings.

\vspace{.2cm}
\noindent 
{{\bf{\large{\underline{Carrollian partition function for bulk gauge fields}}}}}
\vspace{.2cm}

The S-matrix may be computed from a path integral, originally proposed by Arefeva, Faddeev, and Slavnov \cite{Arefeva:1974jv} (see also \cite{Balian:1976vq,Jevicki:1987ax}), with specified asymptotic boundary conditions imposed on the fields. This is in close analogy with the GKP/W dictionary \cite{Gubser:1998bc, Witten:1998qj} of AdS/CFT wherein boundary correlators may be computed from the bulk path integral over fields obeying specified boundary conditions.\footnote{This is contrasted with the BDHM, or ``extrapolate'', dictionary \cite{Banks:1998dd} in which one computes bulk correlation functions before dragging the external legs off to the boundary.} In previous work \cite{Kraus:2024gso, Kim:2023qbl} we have given the prescription for this path integral in the case of scalars and Abelian gauge fields, and we provide a streamlined review of the scalar case in section \ref{Sec: Review}.

In this work we are concerned with the generalization to non-Abelian gauge fields, as an intermediate case between Abelian gauge theory and gravity. In $d=4$, we assume that sufficient matter is present to make the beta function positive, so that gluon scattering  makes sense,  though the matter will only rarely appear explicitly.  We work in Minkowski space, and defining the retarded and advanced times
\eq{inta1}{
    u = t - r,\ \ \ v = t + r
}
we use coordinates $(r, u, \hat x)$ and $(r, v, \hat x)$ near future null infinity, $\scrI^+$, and past null infinity, $\scrI^-$, respectively. In these coordinates, the Minkowski metric is written
\eq{inta2}{
    \dr s^2 &= -\dr u ^2 - 2\dr u \dr r + r^2 \gamma_{AB}\dr \hat x^A \dr \hat x^B\cr
    &= -\dr v^2 - 2\dr v \dr r + r^2 \gamma_{AB}\dr\hat x^A \dr \hat x^B
}
where $\gamma_{AB}$ is the metric of the unit sphere. We give our conventions for the gauge field in appendix \ref{Sec: Conventions}. We then define, suppressing the possible matter content, the path integral for the Carrollian partition function
\eq{inta3}{
    Z[\overline A\subsuper{0A}{a-}, \overline A\subsuper{0A}{a+}] = \int_{\overline A}[\mathcal{D}A\mathcal{D}c\mathcal{D}\overline c]e^{iI[A, \overline A, c, \overline c]}.
}

\begin{figure}[h]
    \centering
    \includegraphics[scale=0.37]{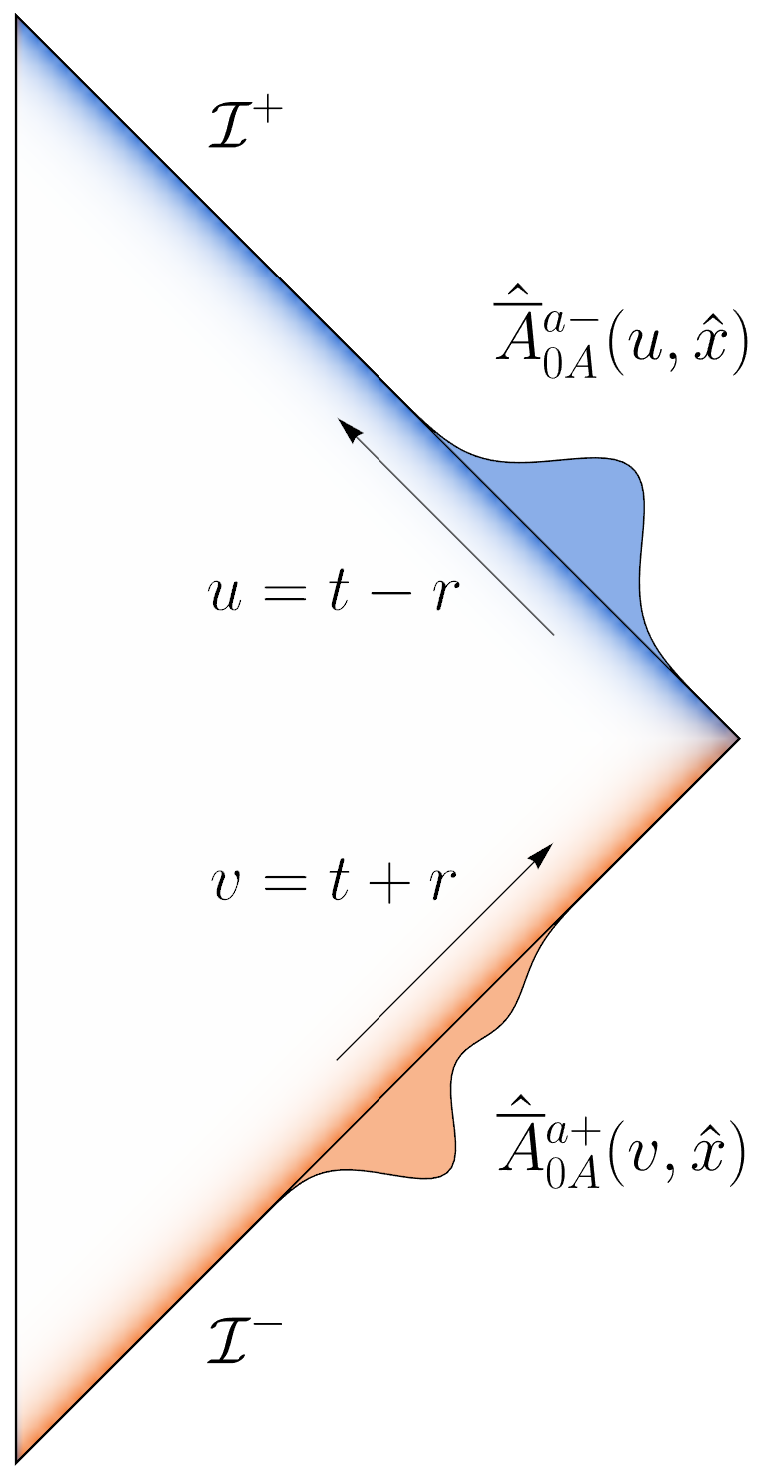}
    \caption{Minkowski Penrose diagram with asymptotic boundary conditions indicated; a pure negative (positive) frequency waveform is specified along $\scrI^+$($\scrI^-$). }
    \label{fig:triangle_with_waveform}
\end{figure}

The boundary conditions here fix the negative (positive) frequency content of the leading behavior of the field near $\scrI^+$ ($\scrI^-$). For example, near $\scrI^+$ this means defining the asymptotic transverse components
\eq{inta4}{
    A\subsuper{0A}{a}(u, \hat x) = \lim_{\substack{r\arrow\infty\\ u \text{ const.}}} A_{A}^a(u, r, \hat x)
}
and fixing the negative frequency components to be $\overline A\subsuper{0A}{a-}(u,\xh)$. The corresponding leading negative/positive frequency component of the ghosts are fixed to zero, equivalent to demanding that \eqref{inta3} compute a bulk transition amplitude between states of vanishing ghost number. The situation is summarized in figure \ref{fig:triangle_with_waveform}.

We must also specify the boundary terms in the action appearing in \eqref{inta3}, the analog of holographic renormalization terms in AdS/CFT.\footnote{As in AdS, these are determined by demanding a good variational principle, a fact we demonstrate in appendix \ref{Sec: LSZ = AFS}. Unlike in AdS, finiteness of the on-shell action is not an important physical requirement,  given the existence  of physically relevant IR divergences. This perspective is more akin to \cite{Papadimitriou:2007sj}.} Computing these boundary terms reveals the importance of certain falloff conditions obeyed by both frequency components of \eqref{inta4} at large $|u|$, i.e. at the bounds of null infinity. We return to these conditions shortly. Since the unfixed frequency content of \eqref{inta4} is fluctuating, we take the view that one cannot impose falloff conditions on it a priori; rather, these falloffs are determined dynamically. As a check, we introduce a toy model in section \ref{Sec: Simplified scattering setup} and perform an explicit check of the falloffs that we assume.

We will be interested in studying the invariance of \eqref{inta3} under large gauge transformations (LGTs)\footnote{The role of LGTs as physical symmetries is clear from the perspective of classical physics, and we recommend \cite{Banados:2016zim,Compere:2018aar} for introductory comments on the mechanism. In appendix \ref{APP: gauge invariance} we offer some remarks instead from the perspective of BRST symmetry within the path integral \eqref{inta3}.}, here meaning transformations
\eq{inta4a1}{
    \delta_{\lambda_-} \overline A\subsuper{0A}{a-}(u \hat x) &= \p_A\lambda^a_-(\hat x) + g\struct{a}{b}{c}\overline A\subsuper{0A}{b-}(u, \hat x)\lambda^c_-(\hat x),\cr
    \delta_{\lambda_+} \overline A\subsuper{0A}{a+}(v \hat x) &= \p_A\lambda^a_+(\hat x) + g\struct{a}{b}{c}\overline A\subsuper{0A}{b+}(v, \hat x)\lambda^c_+(\hat x),
}
for $\lambda_\pm^a(\hat x)$ arbitrary functions on the sphere. These transformations of the data can be extended to gauge transformations in the bulk of the path integral, but the presence of the gauge-fixing term generally makes the invariance of $Z$ non-obvious.

Background field gauge avoids this issue. In this gauge we write $A_\mu^a = \overline A\subsuper{\mu}{a} + a_\mu^a$ where we are free to choose $\overline A{_\mu^a}$ to be any background field configuration obeying the boundary conditions so the fluctuations $a_\mu^a$ have vanishing asymptotic negative (positive) frequency component on $\scrI^+$ ($\scrI^-$).\footnote{For this to hold, $\overline A{_\mu^a}$ must also have vanishing positive (negative) frequency component on $\scrI^+$ ($\scrI^-$).} With this decomposition, one may consider the extension
\eq{inta4a2}{
    \delta_\lambda \overline A\subsuper{\mu}{a} &= \p_\mu \lambda^a + g\struct{a}{b}{c}\overline A{^b_\mu} \lambda^c\cr
    \delta_\lambda a_\mu^a &= g\struct{a}{b}{c}a{^b_\mu} \lambda^c
}
of \eqref{inta4a1} into the bulk, which is not a gauge transformation since the fluctuating field transforms in the adjoint, the affine component of the gauge transformation having been absorbed into the transformation of the background field. It's simple to check that the gauge-fixing and ghost terms are manifestly invariant under \eqref{inta4a2}, and hence the invariance of \eqref{inta3} reduces to invariance of the boundary terms, which is easy to check. This is a notable improvement over the arguments in \cite{Kim:2023qbl}, which worked in Lorenz gauge and required a cancellation between bulk and boundary terms to find invariance.

Since \eqref{inta3} is the Carrollian partition function it is, by definition, the generating functional for Carrollian boundary correlators, as argued in \cite{Kraus:2024gso}. For example, expanding \eqref{inta3} in a power series in the external data, it will contain a term
\eq{inta5}{
    Z[\overline A\subsuper{0A}{a-}, \overline A\subsuper{0A}{a+}] &\supseteq \frac{1}{(2!)^2}\int_{\substack{x_1,x_2\in\scrI^+\\ y_1,y_2\in\scrI^-}}\dr^3 x_1\sqrt{\gamma(x_1)}\dr^3 x_2 \sqrt{\gamma(x_2)}\dr^3 y_1 \sqrt{\gamma(y_1)} \dr^3 y_2 \sqrt{\gamma(y_2)}\cr
    &\!\!\!\!\!\!\!\cdot W_{2,2}^{a_1A_1,a_2A_2;b_1B_1,b_2,B_2}(x_1, x_2; y_1, y_2) \overline A\subsuper{0A_1}{a_1-}(x_1)\overline A\subsuper{0A_2}{a_2-}(x_2)\overline A\subsuper{0B_1}{b_1+}(y_1) \overline A\subsuper{0B_2}{b_2+}(y_2).
}
This term is related to the S-matrix element for $gg\arrow gg$. By directly computing the variations of \eqref{inta3}, we determine the bulk operator conjugate to variations of the boundary data, finding $W$ to be given by correlators of $\sim \p_u a\subsuper{0A}{a+}$  and $\sim \p_u a\subsuper{0A}{a-}$ on $\scrI^+$ and $\scrI^-$, respectively.\footnote{Note that the definite frequency asymptotic behavior of  the background field $\overline{A}{^a_\mu}$ implies that   $\p_u a_{0A}^{a+} = \p_u A_{0a}^{a+}$ on $\Ic^+$, and  $\p_v a_{0A}^{a-} = \p_v A_{0a}^{a-}$ on $\Ic^-$.  } The precise expressions for the conjugate operators are given in \eqref{YM8}. Hence, the boundary correlators $W$ picked out by variations of \eqref{inta3} correspond to what would be called the first $u$-descendant of the Carrollian primary elsewhere in the literature, e.g. in \cite{Donnay:2022aba,Bagchi:2022emh,Donnay:2022wvx,Bagchi:2023fbj,Mason:2023mti,Alday:2024yyj}. If we refer to the operator dual to $\overline A\subsuper{0A}{a\mp}$ in a hypothetical Carrollian dual as $\mathcal{J}_\pm^{aA}$, the boundary correlator in \eqref{inta5} takes the form
\eq{inta6}{
    W_{2,2}^{a_1A_1,a_2A_2;b_1B_1,b_2,B_2}(x_1, x_2; y_1, y_2) = \langle \mathcal{J}^{a_1A_1}_+(x_1) \mathcal{J}^{a_2A_2}_+(x_2) \mathcal{J}^{b_1B_1}_-(y_1)\mathcal{J}^{b_2B_2}_-(y_2)\rangle~,
}
though at the level of our bottom-up approach this may be viewed just as a convenient notation.

An  advantage of the path integral formulation for the Carrollian partition function \eqref{inta3} is that it treats asymptotic symmetries on the same footing as more standard global symmetries, like Lorentz invariance. In both cases, a symmetry induces an action on the asymptotic data $\overline A\subsuper{0A}{a\pm}$ under which the partition function \eqref{inta3} is invariant. Reinterpreting the partition function as the generating functional \eqref{inta5}, the invariance implies Ward identities satisfied by the boundary correlators. It was shown in \cite{Kraus:2024gso} that the soft photon theorem \cite{Low:1958sn,Weinberg:1965nx} and the global Carrollian Ward identities follow in this fashion from large gauge invariance and bulk Lorentz invariances, respectively.

\vspace{.2cm}
\noindent 
{{\bf{\large{\underline{Results}}}}}
\vspace{.2cm}


\vspace{.1cm}
\noindent 
{{\bf{\normalsize{\underline{Disclaimer on regulators}}}}}
\vspace{.1cm}

We now give a description of our results. We are primarily interested in dimension $d = 4$, but the presence of IR divergences make certain arguments delicate at best. For this reason, in section \ref{Sec: Generalization to arbitrary even dimensions} we consider arbitrary (even\footnote{The analysis in odd dimensions is more difficult due to qualitative differences in solutions to the wave equation. See \cite{He:2023bvv} for analysis of the canonical formalism for  Yang-Mills theory in arbitrary dimensions.}) dimensions $d > 4$ where such issues do not arise. Where we state results in $d = 4$, we generally have in mind that one should work in $d > 4$ and continue back to $d = 4$. This continuation procedure is delicate, so we comment where we believe subtleties might arise and provide further discussion in section \ref{Sec: Future directions}, but leave a careful analysis to future work.


\vspace{.1cm}
\noindent 
{{\bf{\normalsize{\underline{Field falloffs}}}}}
\vspace{.1cm}

In the literature, assumptions about the asymptotic behavior of the fields are often imposed to ensure that certain charges converge. For example, the asymptotic field \eqref{inta4} is frequently assumed to fall off faster than $1/u$ as $|u|\arrow\infty$, a requirement which can be schematically understood as the demand that the $\dr u$ integral in the symplectic form $\Omega\sim\int_{\scrI^+}\dr^3 x\sqrt{\gamma} \gamma^{AB}\delta A\subsuper{0A}{a}\wedge \p_u \delta A\subsuper{0B}{a}$ converge. With the boundary conditions in \eqref{inta3}, a significantly weaker condition is sufficient. We observe this condition at the level of the boundary terms in \eqref{inta3}, but the mechanism is simple to understand from the symplectic form, which in terms of our positive/negative frequency split takes the Darboux form $\Omega \sim \int_{\scrI^+}\dr^3 x \sqrt{\gamma}\gamma^{AB}\delta \overline A\subsuper{0A}{a-}\wedge \p_u \delta A\subsuper{0B}{a+}$. Since the positive/negative frequency data are independent, they need not obey the same falloffs.

Specifically, focusing on $\scrI^+$ the fluctuating part of the field $A\subsuper{0A}{a+}$ is allowed to {\em diverge} as $\ln|u|$. To state the conditions on the fixed data, we write $\overline A\subsuper{0A}{a-}(u, \hat x) = \hatline{A}\subsuper{0A}{a-}(u, \hat x) + C_A^{a-}(\hat x)$ where $C_A^- = -\frac{i}{g}e^{-ig\Phi_-}\p_A e^{ig\Phi_-}$ is the flat connection associated with the Goldstone $\Phi^a_-(\hat x)$ of large gauge transformations, which we sometimes refer to as the ``soft'' data on $\scrI^+$. The ``hard'' data $\hatline{A}\subsuper{0A}{a-}$ is allowed to fall off as $1/u$ near the bounds of null infinity. One might suppose that the hard data can be taken to fall off as fast as is necessary but, as we will see in section \ref{Sec: Review}, there is a loss of resolution on the boundary correlators \eqref{inta5} if asymptotic $1/u$ behavior is not allowed.\footnote{This is essentially an issue with how the functional derivative is defined on the space of positive/negative frequency functions. It's easy to see that an issue arises by noting that bump functions cannot be formed in this function space.}

The divergent behavior of the fluctuating field at large $u$ can be understood as a consequence of the soft poles present in momentum space.\footnote{The necessity of this log behavior has recently been noted in work on the log soft theorem, e.g.  \cite{Karan:2025ndk}.} To demonstrate this directly in position space, we describe a simple toy model in section \ref{Sec: Simplified scattering setup} and compute, from position space graphs that we refer to as flat space Witten diagrams, tree amplitudes with this behavior.

We also note that since the conjugate operators are not the fluctuating component of the asymptotic field, but the $u$- or $v$-derivative thereof, the boundary correlators $W$ have the much better asymptotic behavior $1/u$ near the bounds of null infinity. This has been noticed, e.g. in \cite{Mason:2023mti}, where it was observed that the correlators of Carrollian primaries, i.e. the correlators of the fluctuating field component $\sim a\subsuper{0A}{a\pm}$, generally contain IR divergences while the first descendants are well-behaved, at least at tree level.

 In our toy model, we also compute a matrix element of the dynamically determined part of the field strength $F_{AB}|_{\scrI^+}$ to lowest order in the YM coupling. This matrix element explicitly falls off as $1/u$ on $\scrI^+$, so despite the log behavior of the fluctuating field, there are no dynamically generated long range magnetic fields at leading order.


\vspace{.1cm}
\noindent 
{{\bf{\normalsize{\underline{The soft sector}}}}}
\vspace{.1cm}

The soft sector of the theory, defined by variations of the Goldstone, has a number of features made transparent in the present formalism. Firstly, the reader will note that we have not assumed that the Goldstone be antipodally matched,  meaning we have not assumed $\Phi_+^a(\hat x) = \Phi_-^a(-\hat x)$. Rather, the two Goldstones are a priori independent,  hence we write
\eq{inta7}{
    Z = Z[\hatline{A}\subsuper{0A}{a-}(u, \hat x), \Phi\subsuper{-}{a}(\hat x); \hatline{A}\subsuper{0A}{a+}(v, \hat x), \Phi\subsuper{+}{a}(\hat x)].
}
In the bulk, \eqref{inta7} represents a transition amplitude between states labeled in part by the value of the Goldstone, so we should be allowed to specify these states independently, unless the dynamics impose a restriction.

For our purposes, a more useful question is to ask how the operator conjugate to Goldstone variations depends on the choice of initial and final Goldstone value. The contributions to this operator coming from null infinity are independent, but we show in section \ref{Sec: Variations at spatial infinity} that in dimensions $d > 4$ the conjugate operator will receive contributions supported on spatial infinity unless we choose to antipodally match the Goldstone's variations. In brief, if spatial infinity is resolved by a foliation of dS$_3$ slices, the variation of the action at   spatial infinity induced by a Goldstone variation vanishes if the Goldstone obeys the dS$_3$ wave equation. This is the actual condition, also implied by Lorenz gauge for the Abelian theory, used in \cite{Campiglia:2017mua} to demonstrate antipodal matching. In this manner we will not take antipodal matching to be a fundamental principle, but a convenient choice which may or may not be dynamically enforced.

If we make the choice to antipodally match the Goldstone and use only the Goldstone on $\scrI^+$, $\Phi_-^a(\hat x)$, as independent data, we find\footnote{This is strictly only correct to linearized order in the value of the Goldstone. We comment on the generalization to finite Goldstone in appendix \ref{Sec: Charge algebra at finite Goldstone}.}
\eq{inta8}{
    \frac{\delta Z}{\delta \Phi_-^a(\hat x)} = -\frac{1}{2}\hat\nabla_A\left[ \int_{\scrI^+}\dr u \frac{\delta }{\delta\hatline{A}\subsuper{0A}{a-}(u, \hat x)} - \int_{\scrI^-}\dr v \frac{\delta}{\delta \hatline{A}\subsuper{0A}{a+}(v, -\hat x)} \right] Z
}
where $\hat \nabla_A$ is the covariant derivative on the sphere. That is to say, an automatic output of the path integral \eqref{inta3} is that all variations in the soft sector can be replaced by certain variations in the hard sector. This is to be expected given that  in perturbative amplitude calculations in momentum space there is no soft particle line with its own propagator and Feynman rules. There are only hard particles on which we may take a soft limit.

Using \eqref{inta8}, the invariance of \eqref{inta7} under large gauge transformations may be written in infinitesimal form as the Ward identity
\eq{inta9}{
    \left( \hat Q_+^a(\hat x) - \hat Q_-^a(\hat x) \right)Z = 0
}
where the charge, for example on $\scrI^+$, is represented by the differential operator
\eq{inta10}{
    i\hat Q_+^a(\hat x) = \int_{\scrI^+}\dr^3y \sqrt{\gamma}\left[ -\frac{1}{2}\delta_c^a \hat\nabla_{yA}\delta^{(2)}(\hat y - \hat x) + g\struct{a}{b}{c}\hatline{A}\subsuper{0A}{b-}(u, \hat y)\delta^{(2)}(\hat y - \hat x) \right]\frac{\delta}{\delta \hatline{A}\subsuper{0A}{c-}(u, \hat y)}.
}
These charges take the form of vector fields, so their commutator matches the Lie bracket
\eq{inta11}{
    \relax[\hat Q_+^a(\hat x), \hat Q_+^b(\hat y)] = i\struct{c}{a}{b}\delta(\hat x - \hat y)\hat Q^c_+(\hat x)~.
}
Writing the invariance \eqref{inta9} in terms of the boundary correlators as in \eqref{inta5} implies Ward identities (see \eqref{SG4}) which we relate to the standard soft gluon theorem in section \ref{Sec: Soft gluon theorem}. 

The charges \eqref{inta10} may also be written in terms of the operators associated to the fluctuating components of the field as \eqref{SG3b}. Interestingly, the field fall-offs relevant for scattering imply that the charge cannot be written in terms of the total field $\hat A\subsuper{0A}{a} = a\subsuper{0A}{a} + \hatline{A}\subsuper{0A}{a}$; attempting to do so, as in \eqref{SG3e}, reveals a new term that has not previously appeared in the literature.

We demonstrate the relation between the Ward identity \eqref{inta9} and the standard soft gluon theorem in section \ref{Sec: Soft gluon theorem}. In the process we encounter an interesting consequence of the asymptotic $\ln |u|$ behavior  of the fluctuating field, corresponding to asymptotic $1/u$ behavior of the boundary correlators: The integral over null infinity in the first term of \eqref{inta10}, corresponding to the soft leg, is generally only conditionally convergent. Carefully defining this integral in a Lorentz invariant manner produces a relative factor of 2 compared with a naive evaluation, which is crucial for obtaining the correct normalization for the soft gluon theorem. 

It is worth elaborating on this last point.  On the one hand, the Ward identity involves an integral at strictly vanishing $\omega$, since large gauge transformation are constant along the null direction of scri.  On the other hand, we are interested in the soft $\omega \rt 0$ limit of an amplitude.      When relating the Ward identity to the soft limit we therefore  encounter integrals of the form 
\eq{bbb1}{I_1 = \int_{-\infty}^\infty f(u)du~,\quad I_2 = \lim_{\omega \rt 0^+} \int_{-\infty}^\infty f(u)e^{i\omega u}du }
where $f(u)$ is analytic for Im$(u)\geq 0$, and $f(u)\sim {1\over u}$ as $|u|\rt \infty$.   Defining these conditionally convergent integrals as $\int_{-\infty}^\infty (\ldots) du \equiv \lim_{L\rt \infty}\int_{-L}^L (\ldots) du $ it follows that $I_2=2I_1$  (see appendix \rff{Sec: Conditionally convergent integral}), thus providing the factor of $2$ needed to get the correct soft theorem.

To close this discussion of our results, it should be noted that the soft gluon theorem we obtain from the Ward identity \eqref{inta9}, namely \eqref{kk4}, is the one valid for $d > 4$. It is known that  soft theorems in $d = 4$ generally receive loop corrections.\footnote{See \cite{Bern:2014oka} for relevant discussion. Note such corrections also appear in the soft photon theorem when massless charged particles are present in the theory, see e.g. \cite{Ma:2023gir}.  Also, by a ``$d=4$ amplitude", one typically means a dimensionally regularized  amplitude expanded in $\eps$ around $d=4$, with the expansion occurring before the soft limit.}  It would be interesting to explore how the boundary terms, either at null infinity or spatial infinity, should be modified due to the presence of IR divergences and how they either break or deform the large gauge symmetry. We give some further comments in section \ref{Sec: Future directions}, but leave a detailed analysis for future work.

\noindent
{\bf Notation:} Throughout, we work in mostly plus signature, Greek letters $\alpha, \beta,\ldots$ will refer to helicities, capital Latin letters $A, B,\ldots$ refer to directions on the celestial 2-sphere, and lower case Latin letters $a, b, \ldots$, will refer to adjoint indices. The letters $i,j,k\ldots$ will refer to indices in a representation $\mathcal{R}$. We list some additional conventions for the gauge field in appendix \ref{Sec: Conventions}.

\section{Review}
\label{Sec: Review}

We begin by reviewing the essential features of the bulk-boundary dictionary developed in \cite{Kraus:2024gso, Kim:2023qbl}, restricted in this section to the case of a bulk massless real scalar field for simplicity. The presentation here streamlines previous arguments in a way that generalizes well to non-Abelian gauge theories.

\subsection{Path integral, boundary conditions, and action}

We begin by defining the path integral
\eq{r1}{
    Z[\overline\phi{^+_1}, \overline \phi{^-_1}] = \int_{\overline\phi}\D\phi e^{iI[\phi]},
}
introduced in \cite{Arefeva:1974jv}, now interpreted as the Carrollian partition function of a putative Carrollian dual. Defining \eqref{r1} requires that we specify the boundary conditions obeyed by the fields, thereby establishing the integration domain, and the precise action appearing in the integrand. The relevant boundary conditions fix the asymptotic positive and negative frequency content of the fields at null infinity,
\eq{r2}{
    \lim_{x\arrow\scrI^+}r\phi(x) &= (\overline\phi{^-_1}(u, \hat x) + \text{pos. freq.}),\cr
    \lim_{x\arrow\scrI^-}r\phi(x) &= (\overline\phi{^+_1}(v, \hat x) + \text{neg. freq.}).
}
Throughout this work $u$ and $v$ will denote, respectively, the retarded and advanced times
\eq{r3}{
    u = t - r,\ \ \ \ v = t + r
}
in terms of which the Minkowski metric is written
\eq{r4}{
    \dr s^2 &= -\dr u^2 - 2\dr u \dr r + r^2\gamma_{AB}\dr x^A \dr x^B\cr
    &= - \dr v^2 - 2 \dr v \dr r + r^2 \gamma_{AB}\dr x^A \dr x^B.
}
Here $\gamma_{AB}$ is the metric of the unit sphere and we denote points on the sphere by $\hat x$, or later by the stereographic coordinates $(z, \overline z)$ related by
\eq{r5}{
    \hat x = \frac{1}{1 + z \overline z}(z + \overline z, -i(z - \overline z), 1 - z \overline z).
}
It will often be useful to define the null vector
\eq{r5a}{
    n^\mu(\hat x) = (1, \hat x).
}

As in AdS/CFT, the action in \eqref{r1} must supply a good variational principle with respect to our boundary conditions.\footnote{We show that this is equivalent to the demand that \eqref{r1} reproduce the same S-matrix elements computed by Feynman diagrams in appendix \ref{Sec: LSZ = AFS}.} If we write the bulk component of the action in the form
\eq{r6}{
    I_\text{bulk} = \int\dr^4 x \sqrt{-g}\left( \frac{1}{2}\phi\nabla^2\phi - V(\phi) \right)
}
it is straightforward to compute the on-shell variation
\eq{r7}{
    \delta I_\text{bulk} = -(\phi_1, \delta\phi_1)_{\scrI^+} + (\phi_1, \delta\phi_1)_{\scrI^-}
}
where we have introduced the Klein-Gordon-like product
\eq{r8}{
    (A, B) = \frac{1}{2}\int\dr^3x \sqrt{\gamma} (A\p B - \p A B)
}
with the derivative representing $\p_u$ on $\scrI^+$ and $\p_v$ on $\scrI^-$. In the special case of the stereographic coordinates \eqref{r5}, the integration measure is explicitly $\sqrt{\gamma}\dr^3 x = \frac{2\dr u \dr z \dr\overline z}{(1 + z\overline z)^2}$. Using the boundary conditions we split the fields into their positive and negative frequency components so\footnote{Here we drop terms like $(\phi\subsuper{1}{+}, \delta\overline\phi\subsuper{1}{+})_{\scrI^+}$, which vanish assuming the integral \eqref{r8} is convergent. Issues of convergence are more subtle in the case of gauge fields, and will be revisited in section \ref{Path Intgral, boundary conditions, and action}.}
\eq{r9}{
    \delta I_\text{bulk} = &-2(\phi{^+_1}, \delta\overline \phi{^-_1})_{\scrI^+} - \delta(\overline\phi{^-_1}, \phi{^+_1})_{\scrI^+}\cr
    &+2(\phi{^-_1}, \delta\overline \phi{^+_1})_{\scrI^-} + \delta(\overline\phi{^+_1}, \phi{^-_1})_{\scrI^-}.
}

The required boundary terms are therefore
\eq{r10}{
    I_\text{bndy} &= (\overline\phi{^-_1}, \phi{^+_1})_{\scrI^+} - (\overline\phi{^+_1}, \phi{^-_1})_{\scrI^-}\cr
    &= (\overline\phi{^-_1}, \phi{_1})_{\scrI^+} - (\overline\phi{^+_1}, \phi{_1})_{\scrI^-}
}
where the form in the second line makes manifest that the boundary terms are local in the dynamical variable $\phi_1$. 

\subsection{Conjugate operators, falloff conditions, and boundary correlators}

With the complete action in hand, we can ask what operator is conjugate to the boundary conditions \eqref{r2}, or equivalently, what operator is inserted when we take a variation of the boundary data. This is given by\footnote{
The term suppressed in \eqref{r11} proportional to the equations of motion vanishes since the equations of motion always hold inside the path integral. See section \ref{Sec: Future directions} for further comments on the use of this fact.}
\eq{r11}{
    \delta I = -\int_{\scrI^+}\dr^3 x \sqrt{\gamma} ( \phi{^+_1}\p_u \delta\overline\phi{^-_1} - \p_u\phi{^+_1}\delta\overline\phi{^-_1} ) + \int_{\scrI^-}\dr^3 x \sqrt{\gamma}( \phi{^-_1}\p_v \delta\overline\phi{^+_1} - \p_v\phi{^-_1}\delta\overline\phi{^+_1} ).
}
In order to write these terms in the form $\int\dr^3 x \sqrt{\gamma} \mathcal{O}^\mp\delta\overline\phi{^\pm_1}$, we would need to perform an integration by parts. But the resulting boundary terms may be non-zero, or even divergent, depending on the behavior of the fields at the boundaries of scri.

\subsubsection{Falloffs and variational derivatives}

In principle, we could assert that we choose the boundary data $\overline\phi{^\pm_1}$ to fall off sufficiently fast to set all terms at the boundary of $\scrI$ to zero. However, we must remember that these are not completely free functions living on null infinity, rather they are functions of  positive or negative frequency.\footnote{This space of functions is equivalent to the space of functions analytic and bounded in the lower or upper half of the complex $u$-plane, respectively. This is known in mathematics as a Hardy space.} It is productive at this stage to take a small detour to ask how one defines the functional derivative on the space of positive/negative frequency functions.

We normally define the functional derivative of a functional $I$ to be the object $\frac{\delta I}{\delta \phi\subsuper{1}{-}(u)}$, here suppressing the sphere directions and focusing on the negative frequency data at $\scrI^+$, in the expression
\eq{r11a1}{
    \frac{\dr}{\dr \epsilon}I[\overline\phi\subsuper{1}{-}+\epsilon \delta\overline \phi\subsuper{1}{-}]\Big|_{\epsilon = 0} = \int\dr u \frac{\delta I}{\delta \overline\phi\subsuper{1}{-}(u)}\delta \overline\phi\subsuper{1}{-}(u),
}
and we say that the functional derivative is well-defined if it is independent of the choice of function $\delta\overline\phi\subsuper{1}{-}$. Normally one considers the function space to be all (sufficiently smooth) functions on the real line, so one may choose $\delta\overline\phi\subsuper{1}{-} = \delta(u - a)$, so \eqref{r11a1} computes the desired functional derivative. Being somewhat more careful, since the Dirac distribution is not strictly in the space of smooth functions, one should say that \eqref{r11a1} holds for any nascent representation of the Dirac delta, and one can extract the functional derivative from the limit.

In our case, $\delta\overline\phi\subsuper{1}{-}(u)$ is further restricted to be a negative frequency function, and one cannot form a sequence of functions whose limit is the usual Dirac delta within this function space. This is simplest to see from the Poisson kernel representation,
\eq{r11a2}{
    \delta_\epsilon(u) = \frac{1}{2\pi i}\frac{1}{u - i\epsilon} - \frac{1}{2\pi i}\frac{1}{u + i\epsilon}.
}
The first term here is positive frequency, and so this regularized version of the Dirac delta cannot be formed within our function space.

However, it is not strictly necessary to form the distribution \eqref{r11a2}, which acts as the Dirac delta on all smooth functions. The functional derivative $\frac{\delta I}{\delta\overline\phi\subsuper{1}{-}(u)}$ must be a positive frequency function, and so it would suffice to construct a distribution with the Dirac property on the space of positive frequency functions, i.e. a distribution $\delta^-(u - a)$ such that
\eq{r11a3}{
    \int\dr u \delta^-(u - a) f^+(u) = f^+(a)
}
for any positive frequency function $f^+$. Considering \eqref{r11a2}, it's simple to see that by the residue theorem one may choose $\delta^-(u) = -\frac{1}{2\pi i}\frac{1}{u + i\epsilon}$.\footnote{With this choice, the property \eqref{r11a3} holds for any positive frequency $f^+$ which falls to zero, independent of the rate. Constant $f^+$ are a special case where \eqref{r11a3} is conditionally convergent, playing an important role in section \ref{Sec:Sec:Soft gluon theorem}. See appendix \ref{Sec: Conditionally convergent integral} for further comments.}

As a result, we cannot demand our fixed data fall off faster than $1/u$ near the bounds of scri, or we would be unable to form the Dirac distribution, and hence lose the ability to extract correlators from the generating functional \eqref{r1}. We are therefore only allowed to integrate by parts in \eqref{r11} if the fluctuating data blows up slower than linearly near the boundaries. As we will see in section \ref{Sec: Simplified scattering setup}, interactions between massless fields can produce $\ln u$ behavior at large $u$, which indeed satisfies this sublinear growth condition. Indeed, log growth is equivalent to the presence of soft poles, as required by the soft theorem.

\subsubsection{Conjugate operators}

The variation \eqref{r11} may now be written
\eq{r13}{
    \delta I = \int_{\scrI^+}\dr^3x\sqrt{\gamma} (2\p_u\phi{^+_1})\delta\overline\phi{^-_1} + \int_{\scrI^-}\dr^3 x\sqrt{\gamma} (-2\p_v \phi{^-_1})\delta\overline\phi{^+_1}.
}
Hence we may read off the conjugate operators to write
\eq{r14}{
     \frac{\delta Z}{\delta \overline \phi\subsuper{1}{-}(u, \hat x)} &= 2i\langle \overline \phi\subsuper{1}{-}|\p_u \phi\subsuper{1}{+}(u, \hat x) |\overline \phi\subsuper{1}{+}\rangle = \langle {\mathcal{O}}\subsuper{}{+}(u, \hat x)\rangle_{\overline\phi}\cr
     \frac{\delta Z}{\delta \overline \phi\subsuper{1}{+}(v, \hat x)} &= -2i\langle \overline \phi\subsuper{1}{-}|\p_v\phi\subsuper{1}{-}(v, \hat x)|\overline \phi\subsuper{1}{+}\rangle = \langle {\mathcal{O}}\subsuper{}{-}(v, \hat x)\rangle_{\overline\phi}.
}
Here in the middle terms the states $|\overline\phi\subsuper{1}{\pm}\rangle$ indicate the bulk states implementing the boundary conditions \eqref{r2}, and we have suppressed explicit appearance of the S-matrix operator. On the right, we have written these objects as correlators of the dual operator ${\mathcal{O}}\subsuper{}{\pm}$ in the putative dual Carrollian theory with background field $\overline\phi$ turned on, though at the level of our bottom-up analysis this is simply a notational device.

\subsubsection{Boundary correlators}

As explained in \cite{Kraus:2024gso,Kim:2023qbl}, if we write the boundary conditions in Fourier space by
\eq{r15}{
    \phi^-_1(u, \hat x) &= \frac{i}{2(2\pi)^2}\int_0^\infty\dr\omega b^\dag(\omega \hat x) e^{i\omega u}\cr
    \phi^+_1(v, \hat x) &= \frac{i}{2(2\pi)^2}\int_0^\infty\dr\omega b(-\omega \hat x)e^{-i\omega v}
}
then the standard momentum space S-matrix elements are given by
\eq{r16}{
    \langle q_1\ldots q_N| S | p_1\ldots p_M\rangle =\left[ \prod_{n \text{ out}}^N\left( 2\omega_n (2\pi)^3\frac{\delta}{\delta b^\dag(\vec q_n)} \right)\prod_{m \text{ in}}^{M}\left( 2\omega_m(2\pi)^3\frac{\delta}{\delta b(\vec p_m)} \right)Z\right]\Big|_{b=b^\dagger=0}.
}
Hence $Z$ may be written as a formal power series in $b$ and $b^\dag$ whose coefficients are the S-matrix amplitudes including the momentum conserving delta function,
\eq{r17}{
    Z = \sum \frac{1}{N!M!}\int\mathcal{A}(\vec q_1,\ldots,\vec q_N; \vec p_1,\ldots, \vec p_M)\prod_{n\text{ out}}^N \left( \frac{b^\dag(\vec q_n)\dr^3 q_n}{2\omega_n (2\pi)^3} \right) \prod_{m\text{ in}}^M\left( \frac{b(\vec p_m)\dr^3 p_m}{2\omega_m (2\pi)^3} \right).
}

Inverting to write the mode data in terms of the positive and negative frequency functions, we find
\eq{r18}{
    Z = \sum\frac{1}{N!M!}\int_{\substack{x_n\in\scrI^+\\ y_m\in\scrI^-}} W(x_1,\ldots, x_N; y_1,\ldots, y_M)\prod_{n\text{ out}}^N\left(\dr^3 x_n \sqrt{\gamma} \overline\phi{^-_1}(x_n)\right)\prod_{m\text{ in}}^M\left( \dr^3 y_m\sqrt{\gamma}\overline\phi{^+_1}(y_m) \right)
}
where
\eq{r19}{
    W(x_1,\ldots; y_1,\ldots) = \int_0^\infty \prod_{n\text{ out}}^N \left( \frac{\dr\omega_n}{(2\pi)^2}\p_{u_n}e^{-i\omega u_n} \right)\prod_{m\text{ in}}^M \left( \frac{\dr\tilde \omega_m}{(2\pi)^2}(-\p_{v_m})e^{i\tilde\omega v_m} \right) \mathcal{A}(\omega_1\hat x_1,\ldots; -\tilde\omega_{1}\hat y_1, \ldots).
}
The antipodal relation in converting between the Carrollian and momentum space descriptions can be understood as the difference between labeling the direction the particle moves as it falls into the spacetime and the point on the sphere from which it originates.

We refer to the quantity $W$, which in this presentation is manifestly the boundary correlator formed from insertions of the conjugate operators \eqref{r14}, as the boundary correlator. Via the relation \eqref{r19} to momentum space, we can identify the Carrollian primary operators discussed in the literature on Carrollian holography, e.g. \cite{Donnay:2022aba,Donnay:2022wvx,Bagchi:2022emh,Bagchi:2023fbj,Mason:2023mti,Alday:2024yyj}, with the operators for the fluctuating components of our fields, $\phi\subsuper{1}{+}$ on $\scrI^+$ and $\phi\subsuper{1}{-}$ on $\scrI^-$. Our boundary correlators $W$ would then be identified as the correlators of the first $u$-descendants of the Carrollian primaries. Due to the derivative, the boundary correlators $W$ have better IR behavior, as noted in \cite{Mason:2023mti}, and form the basic quantities generated in the path integral presentation of the bulk-boundary dictionary.\footnote{In \cite{Kraus:2024gso} the correlators of Carrollian primaries, there referred to as the boundary correlators $G$, were discussed but only their $u$ and $v$ derivatives, i.e. $W$, appeared directly.}

As in AdS/CFT, boundary correlators may be computed by summing Witten diagrams. Specifically, a Witten diagram in flat space is a bulk Feynman diagram whose external legs have been stripped and replaced with the bulk-boundary propagators
\eq{r20}{
    K_+(x; x') &= \frac{i}{(2\pi)^2}\frac{1}{(u' + n(\hat x')\cdot x - i\epsilon)^2}\cr
    K_-(x; x') &= \frac{i}{(2\pi)^2}\frac{1}{(v' + n(-\hat x')\cdot x + i\epsilon)^2}
}
for outgoing and ingoing legs, respectively. This was shown in \cite{Kraus:2024gso}, and similar flat space Witten diagrams have been studied in \cite{Donnay:2022wvx,Bagchi:2023fbj}. One can check that the bulk-boundary propagators \eqref{r20} are  such that the bulk field
\eq{r21}{
    \phi(x) = \int_{\scrI^+}\dr^3 x' \sqrt{\gamma}K_+(x; x')\overline\phi{^-_1}(x') + \int_{\scrI^-}\dr^3x' \sqrt{\gamma} K_-(x; x')\overline\phi{^+_1}(x')
}
obeys the free wave equation and the boundary conditions \eqref{r2}. Defining the bulk-bulk Green's function (i.e. Feynman propagator in position space) by
\eq{r22}{
    G_F(x; y) = \langle T\phi(x) \phi(y)\rangle = \frac{1}{(2\pi)^2}\frac{1}{(x - y)^2 + i\epsilon}
}
and which obeys $\nabla^2 G_F(x; y) = i\delta^{(4)}(x - y)$, one can relate the bulk-bulk and bulk-boundary propagators by
\eq{r23}{
    K_+(x; x') &= \lim_{\substack{r'\arrow\infty\\u'=\text{const.}}} \phantom{-}2ir'\p_{u'} G_F(x; x'),\cr
    K_-(x; x') &= \lim_{\substack{r'\arrow\infty\\v'=\text{const.}}} -2ir'\p_{v'} G_F(x; x').
}

As a simple example of the graphical rules for constructing flat space Witten diagrams, suppose the potential in \eqref{r6} were $V(\phi) = \frac{\lambda}{3!}\phi^3$. Then the $s$-channel contribution to $W_4$ is
\eq{r24}{
    W^{(s)}_4(x_1, x_2; y_1, y_2) = \begin{tikzpicture}[baseline=(center.base), scale = 2]
        \begin{feynhand}
            \vertex (i+) at (0, 1);
            \vertex (i-) at (0, -1);
            \vertex (i0L) at (-1, 0);
            \vertex (i0R) at (1, 0);
            \propag[plain](i+) to (i0R);
            \propag[plain](i0R) to (i-);
            \propag[plain](i-) to (i0L);
            \propag[plain](i0L) to (i+);
            \vertex (x1) at (-0.5, 0.5);
            \node (x1Label) at (-0.7071-0.1,0.7071-0.05) {$\mathcal{O}^+(x_1)$};
            \vertex (x2) at (0.5, 0.5);
            \node (x2Label) at (0.7071+0.1, 0.7071-0.05) {$\mathcal{O}^+(x_2)$};
            \vertex (y1) at (-0.5, -0.5);
            \node (y1Label) at (-0.7071-0.1, -0.7071) {$\mathcal{O}^-(y_1)$};
            \vertex (y2) at (0.5, -0.5);
            \node (y2Label) at (0.7071+0.1, -0.7071) {$\mathcal{O}^-(y_2)$};
            \node (center) at (0, 0);
            \vertex (z1) at (0, 0.3);
            \node (z1Label) at (0, 0.5) {$z_1$};
            \vertex (z2) at (0, -0.3);
            \node (z2Label) at (0, -0.5) {$z_2$};
            \propag[plain](x1) to (z1);
            \propag[plain](x2) to (z1);
            \propag[plain](z1) to (z2);
            \propag[plain](z2) to (y1);
            \propag[plain](z2) to (y2);
        \end{feynhand}~.
    \end{tikzpicture}
}
The value of this diagram is computed by
\eq{r25}{
    W^{(s)}_4 = (i\lambda)^2\int\dr^4 z_1\dr^4 z_2 K_+(z_1; x_1)K_+(z_1; x_2) G_F(z_1; z_2)K_-(z_2; y_1)K_-(z_2; y_2).
}

\section{Yang-Mills in background field gauge}

We now turn our attention to the main subject of this work, the Carrollian partition function for non-Abelian gauge fields. We outline our conventions in section \ref{Sec: Conventions}.

\subsection{Path Integral, boundary conditions, and action}
\label{Path Intgral, boundary conditions, and action}

Much like the scalar field discussed in section \ref{Sec: Review}, we consider the path integral as a function of the boundary conditions imposed on the fields. In this case, we demand the sphere components of the connection obey
\eq{YM1}{
    \lim_{x\arrow\scrI^+}A_A = (\overline A{_{0A}^-} + \text{pos. freq.}),\cr
    \lim_{x\arrow\scrI^-}A_A = (\overline A{_{0A}^+} + \text{neg. freq.}).
}
We assume that the $r$ and $u$ components fall off faster. Unlike for the scalar field, where we assumed that the boundary data $\overline \phi{^\pm_1}$ falls off at least as $1/u$ near spatial infinity, we allow the boundary data of the gauge field to contain a constant component, namely the large gauge Goldstone. We represent it explicitly by writing\footnote{
We can freely make this decomposition because $\overline A\subsuper{0A}{\pm}$ is free data. We check explicitly in section \ref{Calculation of FAB} that the dynamically determined part of $F_{AB}|_{\scrI^+_-}$ falls to zero in our toy model.
}
\eq{YM2}{
    \overline A{_{0A}^-}(u, \hat x) &= \hatline{A}{^-_{0A}}(u, \hat x) + C_A^-(\hat x)\cr
    \overline A{_{0A}^+}(v, \hat x) &= \hatline{A}{^+_{0A}}(v, \hat x) + C_A^+(\hat x)
}
where now $\hatline{A}{^\pm_{0A}}$ is required to fall off at least as $1/u$, as with the scalar field and $C_A^\pm(\hat x) = -\frac{i}{g}e^{-ig\Phi_\pm}\p_A e^{ig\Phi_\pm}$, so that to linear order in $\Phi_\pm$ we have $C^\pm_A = \p_A\Phi_\pm + \mathcal{O}(\Phi^2_\pm)$, is a flat connection on the sphere.

Throughout this work we will only make use of the case where the Goldstone is linearized, so $C_A^\pm = \p_A\Phi_\pm$, and antipodally matched, meaning $\Phi_+(\hat x) = \Phi_-(-\hat x)$ and we view $\Phi_-(\hat x)$ as the free data. However, we emphasize that neither of these are necessary. In appendix \ref{Sec: Charge algebra at finite Goldstone} we describe some features of working with finite Goldstone.
The antipodal matching of the Goldstone is more interesting since one must be able to compute the transition amplitude between two states of independently chosen Goldstone, $\langle \Phi_-|\Phi_+\rangle$. This is consistent with the calculation in section \ref{Sec: Variations at spatial infinity} demonstrating that, at least for spacetime dimensions $d > 4$, no restriction is placed on the Goldstone by demanding a good variational principle, and so the path integral with non-matched Goldstone is consistent with diagrammatic calculations. The requirement of matching, should it arise at all, must follow as the dynamical statement that such transition amplitudes are proportional to a (field space) delta $\delta(\Phi_-(\hat x) - \Phi_+(-\hat x))$ supported only when the Goldstone is matched.

With the boundary conditions \eqref{YM1} it is natural to shift the integration variable inside the path integral to $A_\mu = \overline A_\mu + a_\mu$ where $\overline A_\mu$ is any fixed configuration obeying the boundary conditions \eqref{YM1}, e.g. a solution to the free wave equation in Lorenz gauge. Given this decomposition, it is natural, and as we will see very useful, to work in background field gauge. Hence if we write the bulk Yang-Mills action as
\eq{YM3}{
    I_\text{bulk} = -\frac{1}{4}\int\dr^4 x \tr F_{\mu\nu}F^{\mu\nu} + I_\text{gf} + I_\text{ghost},
}
we have
\eq{YM4}{
    I_\text{gf} &= -\frac{1}{2}\int\dr^4 x \tr(\overline D_\mu a^\mu)^2\cr
    I_\text{ghost} &= -\int\dr^4 x \overline c_a\left( \overline D^2 \delta^{a}_c + g\struct{a}{b}{c}\overline D^\mu a_\mu^b \right) c^c~,
}
where the background covariant derivative operator is
\eq{YM4a}{ \overline{D}_\mu = \p_\mu -ig \Ab_\mu~.}
We note that the ghost fields are required to obey the same boundary conditions, \eqref{r2}, as the scalar field but with vanishing fixed data.\footnote{This ensures our external states are free of ghosts. We collect some comments on gauge invariance in appendix \ref{APP: gauge invariance}.} With our falloffs, neither the ghost nor gauge-fixing terms produce contributions to the required boundary terms on null infinity, and so the precise form of these terms will not play a large role in this work. The key property of background field gauge, and what makes it particularly useful in studying large gauge symmetry, is that while the gauge-fixing term breaks gauge transformations of the fluctuating field $a_\mu$, the bulk terms of the action are manifestly invariant under gauge transformations acting on the dynamical fields and the background:
\eq{YM5}{
    \overline A_\mu \arrow U^{-1}(\overline A_\mu + \p_\mu)U,\ \ \ a_\mu \arrow U^{-1}a_\mu U,\ \ \ c \arrow U^{-1}cU.
}

To demand a good variational principle subject to the boundary conditions \eqref{YM1}, we vary the action, omitting terms proportional to the equations of motion since they vanish inside the path integral, to find 
\eq{YM6}{
    \delta I_\text{bulk} &= \int_{\scrI^+}\dr^3 x \sqrt{\gamma} \gamma^{AB} \tr \left( \p_u (\hatline{A}{^-_{0A}} + a{^+_{0A}}) \delta (\hatline{A}{^-_{0B}} + C^-_B + a{^+_{0B}}) \right) - (\scrI^-)\cr
    &= \int_{\scrI^+}\dr^3 x \sqrt{\gamma} \gamma^{AB}\tr\Big( \p_u \hatline{A}{^-_{0A}} \delta \hatline{A}{^-_{0B}} + \p_u (\hatline{A}{^-_{0A}} \delta C^-_B) + \p_u \hatline{A}{^-_{0A}} \delta a{^+_{0B}} \cr
    & \quad\quad\quad\quad\quad\quad\quad\quad\quad\quad + \p_u a{^+_{0A}}\delta(\hatline{A}{^-_{0B}} + C^-_B) + \p_u a{^+_{0A}}\delta a{^+_{0B}} \Big)  - (\scrI^-)\cr
}
where the sphere indices are raised and lowered using the unit round sphere metric, $\gamma_{AB}$. 

The term $\tr\p_u \hatline{A}{^-_{0A}} \delta \hatline{A}{^-_{0B}}$ vanishes because it's the integral of two negative frequency functions which fall off at the bounds of scri, and $\tr\p_u (\hatline{A}{^-_{0A}} \delta C^-_B)$ vanishes by our falloff assumptions on $\hatline{A}\subsuper{0A}{a-}$. The third term, $\tr\p_u\hatline{A}\subsuper{0A}{-}\delta a\subsuper{0B}{+}$ must be canceled by a boundary term while $\tr\p_u a{^+_{0A}}\delta(\hatline{A}{^-_{0B}} + C^-_B)$ is compatible with our boundary conditions and defines the conjugate operator to our boundary data.

The final term, $\tr\p_u a{^+_{0A}}\delta a{^+_{0B}}$ naively vanishes for the same reason as the first: it is the integral of two positive frequency functions. However, this is subtle because $a\subsuper{0A}{a+}$ is allowed to diverge as $\ln|u|$ as $|u|\arrow\infty$, and one may check that for $a\subsuper{0A}{a+} = a\subsuper{0A}{a \ln}\ln(u - i\epsilon)$ this term is log divergent. If we suppose that $a\subsuper{0A}{a+} \sim a\subsuper{0A}{a\ln}\ln(u - i\epsilon) + a\subsuper{0A}{a1}\frac{1}{u-i\epsilon} + \cdots$, only the coefficient of the log behavior survives in this term.\footnote{We assume there is no asymptotic constant, i.e. $u^0$, behavior. This behavior is not fixed in our formalism due to the derivatives in \eqref{YM8}.} The term may therefore be canceled by a log divergent  term located at the boundary of $\Ic$  proportional to $\frac{1}{2}(a\subsuper{0A}{a\ln})^2$. We note that this term is a singlet under \eqref{YM5}, so this term does not spoil invariance under gauge transformations of the background. In the presence of higher log divergences of the form $(\ln u)^n$ for $n > 1$, as may be generated by loop corrections\footnote{Loops may induce asymptotic $(\ln u)^n$ behavior. See section \ref{Sec: Future directions} for comments.}, the existence of appropriate boundary terms becomes unclear and is left to future work.

The required boundary terms, suppressing the log divergent term, are therefore
\eq{YM7}{
    I_\text{bndy} &= - \int_{\scrI^+}\dr^3 x\sqrt{\gamma} \gamma^{AB}\tr \p_u\hatline{A}{^-_{0A} a^+_{0B}} + \int_{\scrI^-}\dr^3 x\sqrt{\gamma}\gamma^{AB}\tr \p_u\hatline{A}{^+_{0A} a^-_{0B}}\cr
    &= (\hatline{A}{^{A-}_{0}}, a{_{0A}})_{\scrI^+} - (\hatline{A}{^{A+}_{0}}, a{_{0A}})_{\scrI^-}
}
where in the second line we have used that the hard data goes as $1/u$ while the fluctuating field goes as $\ln u$ near the boundaries of null infinity -- this latter assumption will be justified at tree level in our toy model in section \ref{Sec: Simplified scattering setup}.

\subsection{Conjugate operators and boundary correlators}
\label{Sec: Dual operators and boundary correlators}

From \eqref{YM6} and the variation of \rf{YM7} we can read off the operators conjugate to variations of our boundary data. We see
\eq{YM8}{
    \frac{\delta Z}{\delta \hatline{A}{^{a-}_{0A}}(u, \hat x)} &= 2i\langle \p_u a{^{a+}_{0B}(u, \hat x)} \gamma^{AB}\rangle = \langle \mathcal{J}^{A+}_a(u, \hat x)\rangle,\cr
    \frac{\delta Z}{\delta \hatline{A}{^{a+}_{0A}}(v, \hat x)} &= -2i\langle \p_v a{^{a-}_{0B}(v, \hat x)} \gamma^{AB} \rangle = \langle \mathcal{J}^{A-}_a(v, \hat x)\rangle,
}
where as in \eqref{r14} we have suppressed explicit appearance of the S-matrix, and here also of the states implementing our boundary conditions. On the right, we have written these variations as correlators of the dual currents $\mathcal{J}^{\pm}_{Aa}$ in the putative Carrollian dual.

The new feature in gauge theory is the conjugate operator to the Goldstone. At this stage we specialize to the antipodally matched Goldstone, $\Phi_+(\hat x) = \Phi_-(- \hat x)$, so a variation of $\Phi_-$ induces a variation on $\scrI^-$ as well:\footnote{Note that variations of the Goldstone holding the other data fixed are not gauge transformations, and hence Goldstone variations of $Z$ are non-zero.}
\eq{YM9}{
    \frac{\delta Z}{\delta \Phi^a_-(\hat x)} &= -i\hat \nabla^A \langle \left( \int_{\scrI^+}\dr u \p_u a{^{a+}_{0A}}(u, \hat x) + \int_{\scrI^-}\dr v \p_v a{^{a-}_{0A}}(v, -\hat x) \right) \rangle\cr
    &= -i\hat\nabla^A\langle \int_{\scrI^+}\dr u F_{uA}^{a+}(u, \hat x) + \int_{\scrI^-}\dr v F_{vA}^{a-}(v, -\hat x)\rangle
}
where $\hat\nabla_A$ is the covariant derivative on the sphere. One could have considered variations of the Goldstone which are not antipodally matched, but then the result is not as simple as separating the two terms appearing in \eqref{YM9}. As we show in section \ref{Sec: Variations at spatial infinity} for $d > 4$, the operator conjugate to non-matched Goldstone variations includes a component supported at spatial infinity.

Going further, since the operator conjugate to the Goldstone is formed from integrals of the conjugate operators \eqref{YM8}, one may write
\eq{YM11}{
    \frac{\delta Z}{\delta\Phi_-^a(\hat x)} = -\frac{1}{2}\hat \nabla_A\left[ \left( \int_{\scrI^+}\dr u \frac{\delta Z}{\delta \hatline{A}{^{a-}_{0A}}(u, \hat x)} - \int_{\scrI^-}\dr v \frac{\delta Z}{\delta\hatline{A}{^{a+}_{0A}}(v, -\hat x)} \right) \right].
}
Hence any variation with respect to the soft data can be replaced by variations of the hard data. This relation will be key in obtaining the soft gluon theorem, which in momentum space is also naturally a statement about limits on hard amplitudes --- there is no special ``soft sector'' with dynamics independent of the hard sector. Rather, the so-called soft sector is completely determined by limits taken on the hard sector.

As in the scalar field case, we can relate boundary correlators of the conjugate operators to momentum space amplitudes by expressing the hard boundary data in terms of mode data. In what follows, we will essentially ignore the Goldstone sector of the generating functional, relying on \eqref{YM11} to compute Goldstone variations later. Writing the bulk background field as 
\eq{YM12}{
    \hatline{A}\subsuper{\mu}{a}(x) = \sum_{\alpha =\pm}\int\frac{\dr^3 p}{(2\pi)^3}\frac{1}{2\omega}\left( \epsilon\subsuper{\mu}{*\alpha}(\hat p)a^a_\alpha(\vec p)e^{ip\cdot x} + \epsilon\subsuper{\mu}{\alpha}(\hat p)a^{a\dag}_\alpha(\vec p)e^{-ip\cdot x} \right)
}
The index $\alpha$ labels helicity, and we will henceforth suppress helicity sums.  
We find it useful to use the polarizations 
\eq{YM13}{
    \epsilon\subsuper{\mu}{+}(\hat p) = \frac{1}{\sqrt{2}}(-\overline z, 1, -i, -\overline z),\ \ \ \ \epsilon\subsuper{\mu}{-}(\hat p) = \frac{1}{\sqrt{2}}(-z, 1, i, -z)
}
where the stereographic coordinates $(z, \overline z)$ define the point \eqref{r5} on the sphere. These polarizations are transverse, $n^\mu(\hat p)\epsilon^\pm_\mu(\hat p) = 0$, and have the property $(\epsilon\subsuper{\mu}{+})^* = \epsilon\subsuper{\mu}{-}$. It will be useful to define $\hat \epsilon\subsuper{A}{\pm} = r^{-1} \p_A x^\mu \epsilon\subsuper{\mu}{\pm}$ so
\eq{YM14}{
    \hat\epsilon\subsuper{z}{+} = 0,\ \ \ \hat\epsilon\subsuper{z}{-} = \frac{\sqrt{2}}{1 + z \overline z},\ \ \ \hat\epsilon\subsuper{\overline z}{+} = \frac{\sqrt{2}}{1 + z \overline z},\ \ \ \hat\epsilon\subsuper{\overline z}{-} = 0.
}
Note that while upper and lower sphere indices are differentiated, there is no such distinction in the position of the helicity index. These reduced polarizations have the important property that they provide the conversion between helicity and sphere indices, i.e. they have the property
\eq{YM14a}{
    \hat\epsilon\subsuper{A}{\beta}(\hat p)\hat \epsilon\subsuper{\alpha}{*A}(\hat p) = \delta\subsuper{\alpha}{\beta}
}
and they make a natural appearance in this role when expanding \eqref{YM12} near null infinity,
\eq{YM15}{
    \hatline{A}\subsuper{A0}{a-}(u, \hat x) &= \frac{i}{2(2\pi)^2}\int_0^\infty\dr\omega \hat\epsilon\subsuper{A}{\alpha}(\hat x)a^{a\dag}_\alpha(\omega \hat x)e^{i\omega u}\cr
    \hatline{A}\subsuper{A0}{a+}(v, \hat x) &= \frac{i}{2(2\pi)^2}\int_0^\infty\dr\omega \hat\epsilon\subsuper{A}{*\alpha}(-\hat x)a^a_\alpha(-\omega \hat x)e^{-i\omega v}.
}

The relation between the generating functional and momentum space amplitudes is essentially the same as the scalar case \eqref{r16}, except now the mode data variations carry additional helicity and adjoint indices. When converting $Z$ from a function of the mode data to a function on the hard Carrollian data, our manipulations are performed independently for each leg of the correlator, and hence there is no loss in working out the transformation for a term with a single leg. For each outgoing leg we find,
\eq{YM16}{
    Z^\text{out}_1 &= \int\frac{\dr^3p}{(2\pi)^3}\frac{1}{2\omega}\A^{a\alpha}_\text{out}(\vec p)a^{a\dag}_\alpha(\vec p)\cr
    &= \int_{\scrI^+}\dr^3 x\sqrt{\gamma}\left[ \frac{1}{2\pi}\hat\epsilon\subsuper{\alpha}{*A}(\hat x)\p_u\int_0^\infty \frac{\dr\omega}{2\pi}\A^{a\alpha}_\text{out}(\omega \hat x)e^{-i\omega u} \right] \hatline{A}\subsuper{A0}{a-}(u, \hat x)
}
from which we identify the leg-by-leg relation between the usual momentum space amplitude and the boundary correlator
\eq{YM17}{
    W_\text{out}^{aA}(u, \hat x) = \frac{1}{2\pi}\hat\epsilon\subsuper{\alpha}{*A}(\hat x)\p_u\int_0^\infty\frac{\dr\omega}{2\pi}\A^{a\alpha}_\text{out}(\omega \hat x)e^{-i\omega u}.
}
By essentially the same manipulations, we find the relation for outgoing legs to be
\eq{YM18}{
    W_\text{in}^{aA}(v, \hat x) = \frac{1}{2\pi}\hat\epsilon\subsuper{\alpha}{A}(-\hat x)(-\p_v)\int_0^\infty\frac{\dr\omega}{2\pi}\A_\text{in}^{a\alpha}(\omega, -\hat x)e^{i\omega v}
}
where we have written the energy and angle arguments separately, rather than as $-\omega \hat x$, to emphasize that we are taking the antipodal point on the sphere, not performing some type of analytic continuation in the momentum.

As with the scalar, the boundary correlators $W$ can be computed graphically from Witten diagrams. The bulk-boundary propagators in this case can be deduced by using \eqref{YM15} in \eqref{YM12}. Indeed, one finds that the same field appearing in \eqref{YM12} may be written\footnote{Strictly, the bulk-boundary propagator should be written with a pair of adjoint indices so it contracts against the boundary data as $K\subsuper{\pm b\mu}{aA}\overline A\subsuper{0A}{b-}$. However, these indices are proportional to $\delta_b^a$, so we skip this detail to reduce clutter.}
\eq{YM19}{
    \hatline{A}\subsuper{\mu}{a}(x) = \int_{\scrI^+}\dr^3 x' \sqrt{\gamma} K\subsuper{+\mu}{A}(x; x')\hatline{A}\subsuper{0A}{a-}(x') + \int_{\scrI^-}\dr^3 x' \sqrt{\gamma}K\subsuper{-\mu}{A}(x; x') \hatline{A}\subsuper{0A}{a+}(x')
}
where
\eq{YM20}{
    K\subsuper{+\mu}{A}(x; x') &= \epsilon\subsuper{\mu}{\alpha}(\hat x')\hat\epsilon\subsuper{\alpha}{*A}(\hat x') K_+(x; x'),\cr
    K\subsuper{-\mu}{A}(x; x') &= \epsilon\subsuper{\mu}{*\alpha}(-\hat x')\hat\epsilon\subsuper{\alpha}{A}(-\hat x')K_-(x; x').
}
Here $K_\pm$ are the scalar bulk-boundary propagators given in \eqref{r20}.

In the next section, we describe a toy model in which a boundary correlator can be computed directly. The Fourier transforms \eqref{YM17} and \eqref{YM18} have been applied to compute some examples of boundary correlators from momentum space amplitudes. For example, the Carrollian $n$-point MHV amplitudes may be found in \cite{Mason:2023mti} where our $W$ is referred to as the first $u$-descendant of the Carrollian primary.

\subsection{Simplified scattering setup}
\label{Sec: Simplified scattering setup}

In this section we describe a toy model in which we can directly compute boundary correlators via the flat space Witten diagrams, as opposed to first computing the momentum space amplitudes and then computing their Fourier transforms \eqref{r19} or \eqref{YM17}, \eqref{YM18}. We consider boundary correlators in a theory of a gauge field coupled to a scalar in representation $\mathcal{R}$ of the gauge group,
\eq{SSS1}{
    I = \int\dr^4 x\left( -\frac{1}{4}\tr F^2 - (D_\mu\phi)^\dagger D^\mu \phi \right) - h\phi^\dagger(x_0)\phi(x_0) + I_\text{gf} + I_\text{ghost} + I_\text{bndy}
}
with
\eq{SSS1a}{
    D_\mu\phi^k = \p_\mu \phi^k - igA_\mu^a (t^a\phi)^k~.
}
The action \rf{SSS1}  includes a pointlike interaction localized at the bulk point $x_0$.  This breaks Poincar\'e  invariance, and hence momentum conservation,  allowing for  non-zero on-shell 3-point amplitudes of massless particles. These 3-point amplitudes are simple to compute while illustrating the central points regarding the soft structure of  Carrollian amplitudes. Furthermore, one can think of the amplitudes computed in this model as building blocks for amplitudes computed in a Poincar\'e invariant theory in which $\phi^\dagger(x_0) \phi(x_0)$ is multiplied against other fields and integrated over $x_0$. 

Other authors have found it useful to consider systems which break translation invariance, usually by working on a non-trivial background geometry, e.g. shockwave backgrounds \cite{deGioia:2022fcn,Gonzo:2022tjm,Adamo:2017nia,Adamo:2021rfq}. One may consider this model a toy version of the examples considered elsewhere.

\subsubsection{Two- and three-point boundary correlators}

To lowest order in $g$ and $h$, the Witten diagrams contributing to the boundary correlator with two scalar lines and  a single external gluon are\footnote{Note that there is no diagram of order $g^1 h^0$  since it vanishes by momentum conservation as noted above. }
\eq{SSS2}{
    W^{aAk}_{j}(y_1, y; y_2) = \ \ 
    \begin{tikzpicture}[baseline=(center.base), scale = 2]
        \begin{feynhand}
            \vertex (i+) at (0, 1);
            \vertex (i-) at (0, -1);
            \vertex (i0L) at (-1, 0);
            \vertex (i0R) at (1, 0);
            \propag[plain](i+) to (i0R);
            \propag[plain](i0R) to (i-);
            \propag[plain](i-) to (i0L);
            \propag[plain](i0L) to (i+);
            \vertex (a) at (-0.676777, 0.323223);
            \node (aLabel) at (-0.676777 - 0.24, 0.323223 + 0.22) {$\mathcal{O}^{*+}_j(y_1)$};
            \vertex (b) at (0.5, 0.5);
            \node (bLabel) at (0.5 + 0.27, 0.5 + 0.26) {$\mathcal{J}^{aA+}(y)$};
            \vertex (c) at (0.323223, -0.676777);
            \node (cLabel) at (0.323223 + 0.3, -0.676777 - 0.18) {$\mathcal{O}^{k-}(y_2)$};
            \node[dot] (center) at (0, 0);
            \node (centerLabel) at (0 - 0.2, 0 - 0.1) {$x_0$};
            \vertex (midpoint) at (-0.338389, 0.161612);
            \node (midpointLabel) at (-0.338389, 0.161612 + 0.13) {$x$};
            \propag[plain](center) to (a);
            \propag[gluon](b) to (midpoint);
            \propag[plain](c) to (center);
        \end{feynhand}
    \end{tikzpicture}
    \ +\ 
    \begin{tikzpicture}[baseline=(center.base), scale = 2]
        \begin{feynhand}
            \vertex (i+) at (0, 1);
            \vertex (i-) at (0, -1);
            \vertex (i0L) at (-1, 0);
            \vertex (i0R) at (1, 0);
            \propag[plain](i+) to (i0R);
            \propag[plain](i0R) to (i-);
            \propag[plain](i-) to (i0L);
            \propag[plain](i0L) to (i+);
            \vertex (a) at (-0.676777, 0.323223);
            \node (aLabel) at (-0.676777 - 0.24, 0.323223 + 0.22) {$\mathcal{O}^{*+}_j(y_1)$};
            \vertex (b) at (0.5, 0.5);
            \node (bLabel) at (0.5 + 0.27, 0.5 + 0.26) {$\mathcal{J}^{aA+}(y)$};
            \vertex (c) at (0.323223, -0.676777);
            \node (cLabel) at (0.323223 + 0.3, -0.676777 - 0.18) {$\mathcal{O}^{k-}(y_2)$};
            \node[dot] (center) at (0, 0);
            \node (centerLabel) at (0 - 0.2, 0 - 0.1) {$x_0$};
            \vertex (midpoint) at (0.161612, -0.338389);
            \node (midpointLabel) at (0.161612 - 0.1, -0.338389-0.08) {$x$};
            \propag[plain](center) to (a);
            \propag[gluon](b) to (midpoint);
            \propag[plain](c) to (center);
        \end{feynhand}
    \end{tikzpicture}~.
}
For the scalar we see from the kinetic term that the bulk-bulk and bulk-boundary propagators are
\eq{SSS2a}{
    G_\beta^\alpha(x; y) &= \delta_\beta^\alpha G_F(x; y)\cr
    K_{\pm \beta}^{\alpha}(x; x') &= \delta^\alpha_\beta K_{\pm}(x; x')
}
in terms of the scalar propagators. These diagrams are computed by
\eq{SSS3}{
    W^{aAk}_{j}(y_1, y; y_2) &= \langle \mathcal{O}^{*+}_j(y_1) \mathcal{J}^{aA+}(y)\mathcal{O}^{k-}(y_2)\rangle\cr
    &= -h\int_{M^4}\dr^4 x K_{+\mu}^{A}(x; y)\big[j^{a\mu}(G_F(x; x_0), K_+(x; y_1))\big]_j^k K_{-}(x_0; y_2)\cr
    &\phantom{=} -h\int_{M^4}\dr^4 x K_{+\mu}^{A}(x; y)\big[j^{a\mu}(K_-(x; y_2), G_F(x; x_0))\big]_j^k K_{+}(x_0; y_1)
}
where for any $F_j^k(x),H_j^k(x)$ we have defined
\eq{SSS4}{
    \big[j^{a\mu}(F, H)\big]^k_j &= -ig ( \p^\mu F^k_\ell (t^a)^\ell_i H^i_j - F^k_\ell (t^a)^\ell_i \p^\mu H^i_j )\cr
    &= -ig \big[\p^\mu F t^a H - F t^a \p^\mu H\big]^k_j~.
}

Of course, in the present case all of our propagators are proportional to the identity on their group indices, so the boundary correlators reduces to
\eq{SSS5}{
   & W_{j}^{aAk}(y_1, y; y_2) \cr
   &= igh (t^a)^k_j K_-(x_0; y_2)\int_{M^4}\dr^4x K_{+\mu}^A(x; y)\big( \p^\mu_x G_F(x; x_0) K_+(x; y_1) -  G_F(x; x_0) \p^\mu_x K_+(x; y_1) \big)\cr
   \cr &\ +igh (t^a)^k_j K_+(x_0; y_1)\int_{M^4}\dr^4 x K_{+\mu}^{A}(x; y)\big( \p^\mu_x K_-(x; y_2) G_F(x; x_0) - K_-(x; y_2)\p^\mu_x G_F(x; x_0) \big) \cr
}
where all propagators have been written in terms of their scalar counterparts, \eqref{r20}, \eqref{r22}, and \eqref{YM20}. In particular, we note that the result at this order is nothing but the same 3-point function in scalar QED multiplied by the generator $t_a$.

We perform the bulk integral of \eqref{SSS5} in appendix \ref{Sec:Toy model 3-point function} and quote the result:
\eq{SSS6}{
    & W_{j}^{aAk}(y_1, y; y_2) \cr
    &= \frac{igh}{(2\pi)^2}(t^a)_j^k \hat\epsilon_\alpha^{*A}(\hat y)\left[ -\frac{n(\hat y_1)\cdot \epsilon^\alpha(\hat y)}{n(\hat y)\cdot n(\hat y_1) - i\epsilon} + \frac{n(-\hat y_2)\cdot \epsilon^\alpha(\hat y)}{n(\hat y) \cdot n(-\hat y_2) + i\epsilon} \right] \frac{K_+(x_0; y_1)K_-(x_0; y_2)}{u_y + n(\hat y)\cdot x_0 - i\epsilon} \cr
}
where $y = (u_y, \hat y)$.

Despite coming from such a simple model, \eqref{SSS6} displays the salient features of Carrollian amplitudes important for our considerations. Foremost, we see that the operator conjugate to the gauge field indeed behaves as $1/u$ near the bounds of scri as we claimed should be accounted for in section \ref{Path Intgral, boundary conditions, and action}. This also implies that the unfixed component of the field, $a^+_{0B}$ blows up as $\ln u$ for large $u$. We can also see in this 3-point correlator the basic structure of the soft theorem, with the soft factors clearly present in the square brackets.

Based on the presence of the soft factors, one might rightly guess that the last factor in \eqref{SSS6} should be related to the Carrollian 2-point function. Indeed this is the case, and to order $g^0 h^1$, the 2-point correlator is given by
\eq{SSS7}{
    W_j^k(y_1; y_2) =  \ \ 
    \begin{tikzpicture}[baseline=(center.base), scale = 2]
        \begin{feynhand}
            \vertex (i+) at (0, 1);
            \vertex (i-) at (0, -1);
            \vertex (i0L) at (-1, 0);
            \vertex (i0R) at (1, 0);
            \propag[plain](i+) to (i0R);
            \propag[plain](i0R) to (i-);
            \propag[plain](i-) to (i0L);
            \propag[plain](i0L) to (i+);
            \vertex (a) at (-0.676777, 0.323223);
            \node (aLabel) at (-0.676777 - 0.24, 0.323223 + 0.22) {$\mathcal{O}^{*+}_j(y_1)$};
            \vertex (b) at (0.5, 0.5);
            \vertex (c) at (0.323223, -0.676777);
            \node (cLabel) at (0.323223 + 0.3, -0.676777 - 0.18) {$\mathcal{O}^{k-}(y_2)$};
            \vertex (midpoint) at (-0.338389, 0.161612);
            \propag[plain](c) to (a);
        \end{feynhand}
    \end{tikzpicture}
    \ + \ 
    \begin{tikzpicture}[baseline=(center.base), scale = 2]
        \begin{feynhand}
            \vertex (i+) at (0, 1);
            \vertex (i-) at (0, -1);
            \vertex (i0L) at (-1, 0);
            \vertex (i0R) at (1, 0);
            \propag[plain](i+) to (i0R);
            \propag[plain](i0R) to (i-);
            \propag[plain](i-) to (i0L);
            \propag[plain](i0L) to (i+);
            \vertex (a) at (-0.676777, 0.323223);
            \node (aLabel) at (-0.676777 - 0.24, 0.323223 + 0.22) {$\mathcal{O}^{*+}_j(y_1)$};
            \vertex (b) at (0.5, 0.5);
            \vertex (c) at (0.323223, -0.676777);
            \node (cLabel) at (0.323223 + 0.3, -0.676777 - 0.18) {$\mathcal{O}^{k-}(y_2)$};
            \node[dot] (center) at (0, 0);
            \node (centerLabel) at (0 - 0.2, 0 - 0.1) {$x_0$};
            \vertex (midpoint) at (-0.338389, 0.161612);
            \propag[plain](center) to (a);
            \propag[plain](c) to (center);
        \end{feynhand}
    \end{tikzpicture}
}
The result at this order is
\eq{SSS8}{
    W_j^k(y_1; y_2) &= \langle \mathcal{O}^{*+}_j(y_1) \mathcal{O}^{k-}(y_2)\rangle\cr
    &= \frac{1}{\pi}\delta_j^k\frac{\hat\delta^{(2)}(\hat y_1 + \hat y_2)}{(u_1 - v_2 -i\epsilon)^2} - ih \delta_j^k K_+(x_0; y_1)K_-(x_0; y_2)
}
where the first term, coming from the free theory, was calculated directly from the path integral in \cite{Kraus:2024gso}. We  identify the non-trivial term in \eqref{SSS8} in the 3-point amplitude \eqref{SSS6}.

\subsubsection{Calculation of $F_{AB}|_{\scrI^+_-}$}
\label{Calculation of FAB}

In this simple model we can explicitly compute matrix elements of the field strength and obtain their behavior at the bottom of $\scrI^+$, or the top of $\scrI^-$. Here we focus on the former, computing the matrix element
\eq{FAB1}{
    M_{ABj}^{ak} = \langle y_1,j|F^{a+}_{AB}(y)|y_2,k\rangle
}
where the operator $F_{AB}^{a+}$ is placed on $\scrI^+$ and the states represent a single incoming particle with representation index $k$ at position $y_2\in\scrI^-$, and an outgoing particle with labels $y_1\in\scrI^+$ and $j$.

To lowest non-trivial order in $g$, $F_{AB}^{a+} = \p_A a\subsuper{0B}{a+} - \p_B a\subsuper{0A}{a+}$, and hence using the map \eqref{YM8}, \eqref{FAB1} is equal to
\eq{FAB2}{
    M_{ABj}^{ak} = \frac{1}{2i}\int\dr u \Big[\p_A W\subsuper{Bj}{ak} - \p_B W\subsuper{Aj}{ak}\Big]
}
in terms of the amplitude \eqref{SSS6}, the integral here is indefinite and the angular derivatives act only on the position $y = (u, \hat y)$ of the gluon. To save writing, define
\eq{FAB3}{
    N^{ak}_j \equiv -\frac{1}{2i}\frac{igh}{(2\pi)^2}(t^a)_j^k K_+(x_0; y_1)K_-(x_0; y_2)
}
so we may write
\eq{FAB4}{
    \frac{1}{2i}\int\dr u W_{Aj}^{ak} =  N^{ak}_j \left[ \frac{n(\hat y_1)\cdot \epsilon_\alpha(\hat y)\hat\epsilon_A^{*\alpha}(\hat y)}{n(\hat y)\cdot n(\hat y_1) - i\epsilon} - \frac{n(-\hat y_2)\cdot \epsilon_\alpha(\hat y)\hat\epsilon_A^{*\alpha}(\hat y)}{n(\hat y)\cdot n(-\hat y_2) - i\epsilon} \right]\ln(u_y + n(\hat y)\cdot x_0 + i \epsilon).
}

Now, we are interested in extracting the leading behavior of \eqref{FAB2} as $u \arrow-\infty$. It is simple to check that if $\hat y = \hat y(z, \overline z)$ and $\hat x = \hat x(w, \overline w)$, then
\eq{FAB5}{
    \frac{n(\hat x)\cdot \epsilon_\alpha(\hat y)\hat \epsilon\subsuper{z}{*\alpha}(\hat y)}{n(\hat y)\cdot n(\hat x) - i\epsilon} = \frac{\overline z - \overline w}{(z - w)(\overline z - \overline w) + i\epsilon} = \p_z\ln\big((z-w)(\overline z - \overline w) + i\epsilon \big)
}
where $\epsilon$ has been redefined by a positive multiple, and where the corresponding $\overline z$ component given by $z\leftrightarrow\overline z$ and similarly for $w$. One checks directly that the antisymmetrized sphere derivatives in \eqref{FAB2} annihilate \eqref{FAB5}.

It follows that the only non-trivial contribution to \eqref{FAB2} arises from where the angular derivatives act on the log in \eqref{FAB4}, leading to the result
\eq{FAB6}{
    M\subsuper{ABj}{ak} = N\subsuper{j}{ak}\left[ \frac{n(\hat y_1)\cdot \epsilon_\alpha(\hat y)}{n(\hat y)\cdot n(\hat y_1) - i\epsilon} - \frac{n(-\hat y_2)\cdot \epsilon_\alpha(\hat y)}{n(\hat y)\cdot n(-\hat y_2) - i\epsilon} \right]\frac{\p_A(n(\hat y)\cdot x_0)\hat\epsilon\subsuper{B}{*\alpha}(\hat y) - \p_B(n(\hat y)\cdot x_0)\hat\epsilon\subsuper{A}{*\alpha}(\hat y)}{u_y + n(\hat y)\cdot x_0 + i \epsilon}.
}
Hence the leading fall-off for this matrix element is $1/u$ near $\scrI^+_-$, despite the log growth in the gauge field. However, it should be noted that in \eqref{FAB4} we chose the integration constant to vanish for simplicity. Nothing in our present formalism fixes this choice, nor is there an argument that it cannot contribute to the matrix elements of $F_{AB}^a$. So while the dynamically determined fall-off for the matrix element goes as $1/u$ near $\scrI^+_-$, some additional principle would need to be invoked to rule out $u^0$ behavior.

\subsection{Ward identity of large gauge transformations}
\label{Sec: Ward identity of large gauge transformations}

With the action defined in section \ref{Path Intgral, boundary conditions, and action}, we can ask about the symmetries of the partition function. In particular, we may ask about large gauge transformations, i.e. gauge transformations which shift the Goldstone $\Phi_\pm$. By virtue of working in background field gauge, the bulk part of the action \eqref{YM3} is manifestly invariant under all gauge transformations of the form \rf{YM5},  which shift the background field $\overline A_\mu$ and hence $\Phi_\pm$. The boundary terms \eqref{YM7} are independent of $\Phi$, and hence also invariant under large gauge transformations, implying that the path integral as a whole is invariant under large gauge transformations.

At this stage, there is no a priori reason why we should need to antipodally match the large gauge parameter which shifts $\Phi$. After all, the boundary terms on $\scrI^\pm$ and the bulk terms are all independently invariant, so to this point there is nothing which forces a link between the large gauge transformation on $\scrI^+$ and $\scrI^-$. However, we have so far been cavalier in our treatment of spatial infinity $i^0$. In dimensions $d > 4$, which is the subject of section \ref{Sec: Generalization to arbitrary even dimensions}, one can analyze the behavior of the fields about $i^0$ and find that if we do not antipodally match $\Phi$, and hence restrict to large gauge transformations respecting this matching, the conjugate operator \eqref{YM9} will receive additional contributions from operators at spatial infinity. This is demonstrated in section \ref{Sec: Variations at spatial infinity}. While generating operators at spatial infinity does not necessarily pose a problem, it is not useful for the purpose of extracting a Ward identity. If it is possible to analytically continue these observations in higher dimensions down to $d = 4$, we would be justified in our restriction to antipodally matched LGTs. However, the procedure of analytic continuation runs into  IR divergences in dimensional regularization as $d \rt 4$. This is of course well known, and is related to the loop corrections that can appear in soft theorems in the presence of massless interactions, e.g. \cite{Ma:2023gir}. We comment further on these issues in section \ref{Sec: Future directions}.

We proceed under the assumption that our LGTs are antipodally matched. The invariance of the partition function implies
\eq{ST1}{
    Z[\hatline{A}\subsuper{0A}{+}, \hatline{A}\subsuper{0A}{-}, C_A] = Z[U^{-1}\hatline{A}\subsuper{0A}{+}U,\, U^{-1}\hatline{A}\subsuper{0A}{-}U,\, U^{-1}(C_A + \p_A)U].
}
Infinitesimally, and about $\Phi = 0$, this implies
\eq{ST2}{
    \frac{\delta Z}{\delta \Phi^a(\hat y)} + g\struct{a}{b}{c}\left( \int_{\scrI^+}\!\dr u \hatline{A}\subsuper{0A}{b-}(u, \hat y)\frac{\delta}{\delta \hatline{A}\subsuper{0A}{c-}(u, \hat y)} + \int_{\scrI^-}\!\dr v \hatline{A}\subsuper{0A}{b+}(v, -\hat y)\frac{\delta}{\delta \hatline{A}\subsuper{0A}{c+}(v, -\hat y)} \right) Z = 0.
}
Combining this invariance with the identity \eqref{YM11}, we find the Ward identity
\eq{ST3}{
    (\hat Q^a_+(\hat x) - \hat Q^a_-(\hat x))Z = 0
}
where
\eq{SG3}{
    i\hat Q^a_+(\hat x) &= \phantom{-}\int_{\scrI^+}\dr^3 y \sqrt{\gamma}  \left[ -\frac{1}{2}\delta^a_c \hat\nabla_{yA}\delta^{(2)}(\hat y - \hat x) + g\struct{a}{b}{c}\hatline{A}\subsuper{0A}{b-}(u, \hat y)\delta^{(2)}(\hat y - \hat x) \right] \frac{\delta}{\delta \hatline{A}\subsuper{0A}{c-}(u, \hat y)},\cr
    i\hat Q^a_-(\hat x) &= -\int_{\scrI^-}\dr^3 y\sqrt{\gamma}\left[ -\frac{1}{2}\delta^a_c \hat\nabla_{yA}\delta^{(2)}(\hat y + \hat x) + g\struct{a}{b}{c}\hatline{A}\subsuper{0A}{b+}(v, \hat y) \hat\delta^{(2)}(\hat y + \hat x) \right] \frac{\delta}{\delta \hatline{A}\subsuper{0A}{c+}(v, \hat y)}.
}
where the delta function on the sphere is defined as 
\eq{SG3a}{\int\!d^2y \sqrt{\gamma} \delta^{(2)}(\yh-\xh) =1~.}
The relations \eqref{YM8} may also be used to write the charges \eqref{SG3} directly in terms of the asymptotic field operators. For example,
\eq{SG3b}{
    Q_+^a(\hat x) &= \int_{\scrI^+}\dr u \hat \nabla^A \p_u a\subsuper{0A}{a+}(u, \hat x) + 2g\struct{a}{b}{c} \gamma^{AB}\int_{\scrI^+}\dr u \hatline{A}\subsuper{0A}{b-}(u,\hat x) \p_u a\subsuper{0B}{c+}(u, \hat x).
}
To make contact with the charge described elsewhere in the literature, we can try to write $Q_+^a$ without reference to the positive/negative frequency split. This is simple in the first term since $\p_u C_A = 0$ and $\hatline{A}\subsuper{0A}{a-}$ falls off as $1/u$. For the second term, using the notation,
\eq{SG3c}{
    (A, B)_u = \int_{\scrI^+}\dr u (A\p_u B - \p_u A B),
}
we have\footnote{Here $\hat A\subsuper{0A}{a} \equiv a\subsuper{0A}{a+} + \hatline{A}\subsuper{0A}{a-}$ is the complete asymptotic field with the Goldstone $C_A^a$ removed.} ,
\eq{SG3d}{
    2\struct{a}{b}{c}(\hatline{A}\subsuper{0A}{b-}, a\subsuper{0B}{c+})_u = \struct{a}{b}{c}\Big[(\hat A\subsuper{0A}{b}, \hat A\subsuper{0B}{c})_u - (\hatline{A}\subsuper{0A}{b-}, \hatline{A}\subsuper{-B}{c-})_u - (a\subsuper{0A}{b+}, a\subsuper{0B}{c+})_u \Big].
}
The second term on the right vanishes since it's the integral of two negative frequency functions which fall off at large $u$. The third term is more subtle, similar to the  boundary term discussed below \eqref{YM6}, and may not vanish.

The charge may therefore be written
\eq{SG3e}{
    Q_+^a(\hat x) = \int_{\scrI^+}\dr u \hat\nabla^A\p_u A\subsuper{0A}{a} + \frac{1}{2}g\struct{a}{b}{c}\gamma^{AB}\Big[ (\hat A\subsuper{0A}{b}, \hat A\subsuper{0B}{c})_u - (a\subsuper{0A}{b+}, a\subsuper{0B}{c+})_u \Big].
}
The final term, involving the product of two fluctuating fields at coincident points on the sphere, has not previously appeared in the literature because it would vanish if we demanded $\hat A\subsuper{0A}{a}(u, \hat x)$ fall off as $1/u$ for $|u|\arrow\infty$ as is sometimes assumed;  however,  we have seen that such a falloff assumption  is generally invalid for the field configurations that arise in scattering.

The matrix elements of this final term are difficult to compute because they receive contributions from collinear gluons, but we can compute the contribution due to the boundary correlator \eqref{SSS6} and find it to be non-zero. However, we emphasize that all of  this is an issue with writing the charge in terms of the field configuration without reference to the positive/negative frequency split. The charge relevant in this work, \eqref{SG3b}, contains no such collinear issues.

Though in the present work we have primarily considered pure Yang-Mills theory, the inclusion of  matter  is straightforward. One checks that the boundary terms of the matter sector do not violate the large gauge symmetry  so the modified Ward identity is again derived from the statement of invariance as in \eqref{ST1}. For example, if we included a scalar field transforming in some representation $\mathcal{R}$ with generators $(t^a)_j^k$, we would find the additional contributions
\eq{ST5}{
    +ig\left( \int_{\scrI^+}\dr u (t^a)_j^k ~ \overline\phi\subsuper{1}{j-}(u, \hat y)\frac{\delta}{\delta \overline \phi\subsuper{1}{k-}(u, \hat y)} + \int_{\scrI^-}\dr v (t^a)_j^k ~ \overline\phi\subsuper{1}{j+}(v, -\hat y) \frac{\delta}{\delta \overline\phi\subsuper{1}{k+}(v, -\hat y)} \right)Z
}
on the LHS of \eqref{ST2}. Corresponding contributions would then be added to the operators \eqref{SG3}.

\subsection{Soft gluon theorem}
\label{Sec: Soft gluon theorem}

To connect the Ward identity \eqref{ST3} to the standard soft theorem,\footnote{\label{ft1}We restrict attention to the leading soft gluon theorem. See section \ref{Sec: Future directions} for some comments on a possible approach to the subleading case.} we use that $Z$ is the generating functional of boundary correlators as we did in section \ref{Sec: Dual operators and boundary correlators}. Since the differential operators \eqref{SG3} are first order, it is sufficient to understand how they act on a single leg. For an outgoing leg, and suppressing dependence on other legs, we have
\eq{SG1}{
    Z_1^\text{out} = \int_{\scrI^+}\dr^3 x \sqrt{\gamma} W_\text{out}^{aA}(u, \hat x) \hatline{A}\subsuper{0A}{a-}(u, \hat x),
}
and so
\eq{SG2}{
    i\hat Q^a_+(\hat y) Z_1^\text{out} = \frac{1}{2}\int_{\scrI^+}\dr u \hat\nabla_AW_\text{out}^{aA}(u, \hat y) + \int_{\scrI^+}\dr^3 x \sqrt{\gamma}\left( g\struct{a}{b}{c}\delta^{(2)}(\hat y - \hat x) W^{cA}_\text{out}(u, \hat x) \right) \hatline{A}\subsuper{0A}{b-}(u, \hat x).
}

So we find two terms, one originating from the variation of the Goldstone, which by \eqref{YM11} is equivalent to the affine part of the gauge field's transformation, and the other originating from the adjoint part of the gauge field's transformation. Since these two terms now come with a different number of factors of $\hatline{A}$, they  mix with other terms that we have omitted in writing \eqref{SG1}. The first term in \eqref{SG2} would need to cancel against the adjoint part of $\hat Q Z_0$ (were there any), and the second term would need to cancel against the Goldstone part of $\hat Q Z_2$. One must be a little careful when including terms with both ingoing and outgoing legs since both $Z_1^\text{out}$ and $Z_1^\text{in}$ will produce a term with the same number of $\hatline{A}\subsuper{A0}{\text{in/out}}$ legs from the Goldstone part of the transformation. The complete result for the Ward identity associated to large gauge invariance may therefore be written
\eq{SG4}{
    -&\frac{1}{2}\hat\nabla_{yA}\Bigg[ \int_{\scrI^+}\dr^3 y'\sqrt{\gamma} \delta^{(2)}(\hat y - \hat y')W_\text{out}^{aA;b_1B_1;\ldots}(y', x_1,\ldots) \cr& \quad\quad\quad \quad \quad  + \int_{\scrI^-}\dr^3 y'\sqrt{\gamma} \delta^{(2)}(\hat y + \hat y') W_\text{in}^{aA;b_1B_1;\ldots}(y', x_1,\ldots) \Bigg]\cr
    &= g \sum_{n=1} \struct{a}{b_n}{c} \delta^{(2)}(\hat y - \hat\eta_n\hat x_n) W^{b_1B_1;\ldots;cB_n;\ldots}(x_1,\ldots)
}
where on the first line the boundary correlators are marked by whether the soft gluon leg was ingoing or outgoing and on the second line we have defined
\eq{SG5}{
    \hat\eta_n\hat x_n = \left\{ \begin{array}{ll}
        \phantom{-}\hat x_n, & n\text{ out} \\
        -\hat x_n, & n\text{ in}
    \end{array} \right.~.
}
In the following subsections, we discuss how this Ward identity relates to the standard soft gluon theorem.

\subsubsection{Additional requirements}
\label{Sec: Additional requirements}

To reproduce the soft theorem from the Ward identity \eqref{SG4}, we need to address two points:
\begin{itemize}
    \item[1)] The LHS of \eqref{SG4} involves the average of an incoming and outgoing soft gluon, whereas the standard soft theorem involves only one of the two (and the RHS does not depend on the choice).
    \item[2)] The divergence on the sphere on the LHS of \eqref{SG4} must be inverted so the angular component $A$, equivalent to the helicity $\alpha$ under the map \eqref{YM14a}, can be free. But naively \eqref{SG4} only determines the soft behavior up to some divergence-free vector field.
\end{itemize}
It's therefore necessary to argue that\footnote{In the literature this is the statement that the so-called soft photon/gluon operator, $N_A^a(\hat x) \sim \int_{\scrI^\pm}F_{uA}^a(u, \hat x)$, is antipodally matched \cite{He:2014cra}.}
\eq{AR1}{
    \int_{\scrI^+}\dr^3 y' \sqrt{\gamma} \delta^{(2)}(\hat y - \hat y') W^{aA;b_1B_1;\ldots}_{\text{out}}(y', x_1,\ldots) = \int_{\scrI^-}\dr^3y' \sqrt{\gamma} \delta^{(2)}(\hat y + \hat y')W^{aA;b_1B_1;\ldots}_\text{in}(y', x_1,\ldots)
}
and that both sides of \eqref{AR1} are curl-free on the sphere.

Both of the requirements (1) and (2) have appeared in the existing literature on asymptotic symmetries, for example \cite{Strominger:2017zoo}. There have also been some studies of what one can deduce when (2) is relaxed, this situation usually relating to the presence of magnetic monopoles \cite{Strominger:2015bla}. It would be interesting to understand these arguments directly in the present Carrollian formalism, and in particular the extent to which the equations of motion and the asymptotics near spatial infinity can be used to establish these requirements. However, we leave such investigations to future work.

\subsubsection{Soft gluon theorem}
\label{Sec:Sec:Soft gluon theorem}

With the considerations of the previous subsection in mind, we are free to write the Ward identity \eqref{SG4} in the form
\eq{SG7}{
    \int_{\scrI^+}\dr u W\subsuper{\text{out}}{aA;b_1,B_1;\ldots}((u, \hat y), x_1,\ldots) = -g\sum_{n=1}\struct{a}{b_n}{c}G^A(\hat y, \hat\eta_n\hat x_n)W^{b_1B_1;\ldots;c B_n;\ldots}(x_1,\ldots)
}
where $G^A(\hat y, \hat\eta_n\hat x_n)$ is the vector Green's function on the sphere satisfying  
\eq{SG8}{
   \hat\nabla_{yA} G^A(\hat y, \hat x) = {1\over \sqrt{\gamma(\yh)}} \p_{yA}(\sqrt{\gamma(\hat y)}G^A(\hat y, \hat x)) = \delta^{(2)}(\hat y - \hat x)
}
In the stereographic coordinates $(z, \overline z)$ defined by $\hat x = \frac{1}{1 + z\overline z}(z + \overline z, -i(z - \overline z), 1 - z\overline z)$ this Green's function is\footnote{This follows from $\p_z {1\over \zb}=\p_{\zb}{1\over z} =2\pi \sqrt{\gamma} \delta^{(2)}(z, \overline z)$, where $\delta^{(2)}(z,\zb)$ is  the covariant delta function \rf{SG3a}.  }
\eq{SG9}{
    G^z(\hat y(z,\overline z); \hat x(w, \overline w)) &= \frac{1}{4\pi}\frac{1}{\sqrt{\gamma(\hat y)}}\frac{1}{\overline z - \overline w},\cr
    G^{\overline z}(\hat y(z,\overline z); \hat x(w, \overline w)) &= \frac{1}{4\pi}\frac{1}{\sqrt{\gamma(\hat y)}}\frac{1}{z - w}.
}
These can be written compactly without reference to any choice of coordinates on the sphere as
\eq{SG9a1}{
    G^A(\hat y; \hat x) = \frac{1}{4\pi} \frac{\hat \epsilon^{*A}_\alpha(\hat y)\epsilon^\alpha(\hat y) \cdot n(\hat x)}{n(\hat y) \cdot n(\hat x)}.
}
Hence the Ward identity for an outgoing gluon takes the form
\eq{SG10}{
    \int_{\scrI^+}\dr u W\subsuper{\text{out}}{aA;b_1B_1;\ldots}((u, \hat y), x_1,\ldots) = -\frac{g}{4\pi}\sum_{n = 1}\struct{a}{b_n}{c} \frac{\hat\epsilon^{*A}_\alpha(\hat y) \epsilon^\alpha(\hat y)\cdot n(\hat\eta_n\hat x_n)}{n(\hat y)\cdot n(\hat\eta_n\hat x_n)}W^{b_1B_1;\ldots;cB_n;\ldots}(x_1, \ldots).
}
It is now straightforward to show that the Ward identity in the form \eqref{SG10} implies the standard momentum space statement of the soft theorem.

We are interested in extracting the residue of the soft pole,\footnote{Since the Fourier transforms \eqref{YM17} and \eqref{YM18} commute with the soft factor, there is no loss in leaving the hard legs in their Carrollian description.}
\eq{kk1}{ \lim_{\omega\arrow 0^+}\omega \mathcal{A}_\text{out}^{\alpha;b_1B_1;\ldots}(\omega \hat y; x_1,\ldots) & = 2\pi i \hat{\eps}_A^\alpha(\hat{y})    \lim_{\omega\arrow 0^+}\int^\infty _{-\infty}\! du W\subsuper{\text{out}}{aA;b_1B_1;\ldots}((u, \hat y), x_1,\ldots)  e^{i\omega u}~,   }
where we used \rf{YM17} and  \rf{YM14a}.  We then note that \rf{SG10} appears to give us an expression for the integral appearing in \rf{kk1}.  However, naive substitution into \rf{kk1} gives a result for the soft theorem that is off by a factor of $2$.  Proceeding more carefully, we first note that the integral in \rf{kk1} is only conditionally convergent since  $ W \sim 1/u$ at large $u$, and so the limit $\omega\arrow 0^+$ need not commute with the $\dr u$ integration.  We handle this by defining such integrals as
\eq{kk2}{ \int_{-\infty}^\infty  \! du  f(u) \equiv \lim_{L \rt \infty} \int_{-L}^L \! du f(u)~.  }
Given this definition and the assumption that $f(u) $ is analytic in the lower half $u$-plane including the real axis, as expected for a positive frequency function, and behaves as $f(u) \sim {1\over u} $ as $|u|\rt \infty$,   it follows that 
\eq{kk3}{  \lim_{\omega\arrow 0^+} \int_{-\infty}^\infty \! du f(u) e^{i\omega u} = 2 \int_{-\infty}^\infty \! du f(u)~,   }
This result is  proven in appendix \rff{Sec: Conditionally convergent integral}.  Applying this result to \rf{kk1} and \rf{SG10} gives 
\eq{kk4}{ \lim_{\omega\arrow 0^+}\omega \mathcal{A}_\text{out}^{\alpha;b_1B_1;\ldots}(\omega \hat y; x_1,\ldots) 
& =  -ig\sum_{n=1}\struct{a}{b_n}{c}\frac{\epsilon^\alpha(\hat y)\cdot n(\hat\eta_n\hat x_n)}{n(\hat y)\cdot n(\hat\eta_n \hat x_n)}W^{b_1B_1; \ldots; cB_n;\ldots}(x_1, \ldots)}
which we identify as the standard form of the momentum space soft theorem. 

The factor of $2$  discussed above is related to subtle factors of $2$ appearing  in the Dirac brackets of soft modes \cite{He:2014laa,He:2014cra}  which also needed to be taken into account to arrive at the correct soft theorem.   In our argument above, we defined the integrals via a symmetric limiting procedure.  More generally, we could have written $\lim_{L\rt \infty} \int_{u_{\rm min}(L)}^{u_{\rm max}(L)} du f(u)   $,  where the integration limits go to infinity such that  $\lim_{L\rt \infty} {u_{\rm max}(L) \over u_{\rm min}(L)  } =-1 $.    Importantly, this condition is Poincar\'e invariant, recalling that Poincar\'e transformations act on $\Ic^+$ as $u\rt \lambda u + b$.

It is worth commenting on how the soft factor in \rf{kk4} keeps track of  relative  signs associated to the hard gluons. Suppressing the $A$-type indices, and denoting the soft factors as $S_k$, the RHS of \rf{kk4} can be written as $\sum_{n=1} f^a_{~b_n c} S_n W^{b_1,\ldots b_{n-1}, c, b_{n+1}, \ldots}$.  Invariance under global gauge transformations implies that this must give zero upon setting $S_n=1$, which implies that $W^{b_1, b_2, \ldots }$ is an invariant tensor. 
To illustrate, consider the case of  two hard gluons, where the invariant tensor is $W^{b_1 b_2} = \delta^{b_1 b_2}$.   This gives $\sum_{n=1} f^a_{~b_n c} S_n W^{b_1,\ldots b_{n-1}, c, b_{n+1}, \ldots} = f^a_{~b_1b_2} (S_1 -S_2)$.  For a clear statement of the various signs appearing in the non-Abelian soft theorem, see e.g. \cite{Catani:2000pi}.

\subsubsection{Including generic representation matter}

It is simple to include generic representation matter in our analysis. As commented in section \ref{Sec: Ward identity of large gauge transformations}, in the presence of matter in a representation $\mathcal{R}$, the identity \eqref{ST2} would acquire an additional contribution on the RHS of the form \eqref{ST5}. We could follow the analysis of section \ref{Sec: Soft gluon theorem} and track the new terms through to find an additional contribution on the RHS of \eqref{SG10}, noting that standard matter will not alter the arguments given in section \ref{Sec: Additional requirements}.

However, a faster approach would be to first recall that $\struct{a}{b}{c} = i(t^\text{adj}_a)^b_c$ and so \eqref{SG10} may be written
\eq{SGR1}{
    \int_{\scrI^+}\dr u W^{aA;b_1B_1;\ldots}((u,\hat y), x_1, \ldots) = -\frac{ig}{4\pi}\sum_{n=1}(t^\text{adj}_{a})_c^{b_n} G^A(\hat y, \hat\eta_n \hat x_n)W^{b_1B_1; \ldots;cB_n;\ldots}(x_1, \ldots).
}
It follows that a contribution on the RHS from a particle in representation $\mathcal{R}$ would simply be the same as that of the gluon, but where the group label $b_n$ is replaced by a label $k_n$ in the representation $\mathcal{R}$, and the adjoint generator $(t^\text{adj}_{a})_c^{b_n}$ is replaced by the appropriate generator $(t^{\mathcal{R}}_a)^{k_n}_c$, $c$ now interpreted as an index in the representation $\mathcal{R}$.

\subsubsection{Position space check}
\label{Sec:Positions space check}

The regularization prescription used in section \ref{Sec:Sec:Soft gluon theorem} to find the momentum space soft theorem might, at first, seem ad hoc. In this section, we use the explicit boundary correlators \eqref{SSS6} and \eqref{SSS8} to check the large gauge Ward identity.

We begin by writing \eqref{SSS6} in terms of \eqref{SSS8},\footnote{Note that the free term in \eqref{SSS8} makes no contribution here. Due to the $\delta^{(2)}(\hat x_1 + \hat x_2)$, the soft factors combine into a Dirac delta proportional to $\delta^{(2)}(\hat x_1 - \hat x_2)$. This product of deltas has vanishing support. We see that the $i\epsilon$'s appearing in \eqref{SSS6} are critical to giving the correct definition of the amplitude.}
\eq{PSC1}{
    W^{aAk}_j(x_1, y; x_2) = -\frac{g/(4\pi)^2}{u + n(\hat y)\cdot x_0 - i\epsilon}\Big[ (-t^a)^\ell_j &\frac{\hat \epsilon^{*A}_\alpha(\hat y)\epsilon^\alpha(\hat y)\cdot n(\hat x_1)}{n(\hat y)\cdot n(\hat x_1) - i\epsilon}W_\ell^k(x_1; x_2)\cr
    + (t^a)_\ell^k &\frac{\hat\epsilon^{*A}_\alpha(\hat y)\epsilon^\alpha(\hat y)\cdot n(-\hat x_2)}{n(\hat y)\cdot n(-\hat x_2) + i\epsilon} W^\ell_j(x_1; x_2) \Big].
}
Using the definition \eqref{kk2} the $\dr u$ integral is computed to be
\eq{PSC2}{
    I = \int\frac{\dr u}{u + n(\hat y)\cdot x_0 - i\epsilon} = i\pi.
}
Hence, we find
\eq{PSC4}{
    \int_{\scrI^+}\dr u W^{aAk}_j(x_1, y; x_2) = -\frac{ig}{4\pi} \Big[ (-t^a)^\ell_j &\frac{\hat \epsilon^{*A}_\alpha(\hat y)\epsilon^\alpha(\hat y)\cdot n(\hat x_1)}{n(\hat y)\cdot n(\hat x_1) - i\epsilon}W_\ell^k(x_1; x_2)\cr
    + (t^a)_\ell^k &\frac{\hat\epsilon^{*A}_\alpha(\hat y)\epsilon^\alpha(\hat y)\cdot n(-\hat x_2)}{n(\hat y)\cdot n(-\hat x_2) + i\epsilon} W^\ell_j(x_1; x_2) \Big].
}
We now identify this as precisely the form \eqref{SGR1} of the Ward identity relevant in the presence of charged matter fields. We note that the minus sign in $(-t^a)^\ell_j $ is the generalization of the relative minus sign associated to incoming versus outgoing charged particles in the soft photon theorem.

\subsection{Charge algebra}
\label{Sec:Charge algebra}

Since we have access to the complete charges \eqref{SG3}, it is straightforward to compute directly the algebra obeyed by the charges. For example,
\eq{CA1}{
    \relax [\hat Q^a_+(\hat x), \hat Q^b_+(\hat y)] = i\struct{c}{a}{b}\delta^{(2)}(\hat x - \hat y) \hat Q_+^c(\hat x).
}
This is the expected algebra for generators of large gauge transformations with no central extension. We may also note that since the expressions \eqref{SG3} are first order differential operators which are at most linear in the fields, the only ordering ambiguity amounts to an additive constant which, upon recalling \eqref{ST3}, is removable.

The charges, and their algebra, can also be checked for the case of finite Goldstone. The calculation is more involved and no new features arise, but for sake of completeness we include the details in appendix \ref{Sec: Charge algebra at finite Goldstone}.

\section{Generalization to arbitrary even dimension}
\label{Sec: Generalization to arbitrary even dimensions}

In this section we discuss the extension of our arguments to dimensions $d>4$.    Besides being a natural extension from a theoretical perspective, another reason for considering $d>4$ is that it is simpler in important respects than $d=4$.  First, due to the absence of infrared divergences the S-matrix exists and our Ward identities hold at loop level.   Relatedly, the validity of the Ward identity relies on the absence of symmetry violating boundary terms at spatial infinity.  This is tricky to establish in $d=4$ due to the relatively slow field falloffs (for example it is not sufficient to consider only quadratic terms in the action), while it is straightforward in $d>4$, as we will see.  More precisely, this holds provided we demand that the  Goldstone mode\footnote{More precisely, its variation.}  obeys $\nabla^2 \Phi=0$ at spatial infinity, a condition which imposes the antipodal identification between $\Phi$ at $\Ic^+$ and $\Ic^-$.    This antipodal identification is a key feature needed to obtain the soft theorem. 

\subsection{Even versus odd $d$}

As has been discussed in prior work\cite{Kapec:2014zla,He:2019jjk,He:2019pll,He:2023bvv}, there are significant differences for even versus odd $d$. From our perspective the issue is that the radiative falloff of the gauge field near $\Ic$ is
\eq{zz1}{ A_A \sim r^{-{d-4\over 2}},}
 while the Goldstone mode $\Phi$ obeying $\nabla^2 \Phi=0$ near $\Ic$ has the expansion
 \eq{zz2}{ \Phi(x) = \sum_{n=0}^\infty r^{-n} \Phi_n(u,x^A)~.}
Our derivation of the Ward identity involves relating a shift of the Goldstone mode to a shift of the radiative fields, but this is not possible for odd $d$ due to the integer versus half-integer mismatch in the power of $r$ falloffs.    For this reason, we will restrict to even $d$ in the following, though it would be interesting to better understand the odd $d$ case as well.  \cite{He:2019pll,He:2019jjk,He:2023bvv} handle both even and odd $d$ in a different approach, not related to invariance of a boundary action.

\subsection{Action}

We start by writing down the obvious extension of the $d=4$ action to general $d$.  In particular the boundary terms on $\Ic$ are as in \rf{YM7},
\eq{zz3}{
    I_\text{bndy} &= - \int_{\scrI^+}\dr^{d-1} x\sqrt{\gamma} \gamma^{AB}\tr \p_u\hatline{A}{^-_{0A} a^+_{0B}} + \int_{\scrI^-}\dr^{d-1} x\sqrt{\gamma}  \gamma^{AB}\tr \p_u\hatline{A}{^+_{0A} a^-_{0B}}\cr
    &= (\hatline{A}{^{A-}_{0}}, a{^+_{0A}})_{\scrI^+} - (\hatline{A}{^{A+}_{0}}, a{^-_{0A}})_{\scrI^-}
}
Also, the variation of the action with respect to the Goldstone field is as in \rf{YM9},
\eq{zz4}{ \delta I = \int_{\Ic^+} d^{d-1}x \sqrt{\gamma} \lim_{r\rt \infty}  r^{d-4}  \nabh^A \p_u a^+_A \delta \Phi + (\Ic^-) }
where the factor of $r^{d-4}$ arises from the sphere measure on a fixed $r$ surface.  An immediate issue that arises is that if we assume the falloffs \rf{zz1} and \rf{zz2} then we seem to find terms in \rf{zz4} that diverge with $r$.   To resolve this issue, consider the terms involving the leading part of $a^+_A$, which we write as 
\eq{zz5}{  a^+_A(x) = r^{-{d-4\over 2}} a^+_{0A}(u,\xh) + \ldots }
where we chose the $0$ subscript to make formulas look similar to the $d=4$ case.  The key point is that the large $u$ behavior of $ a^+_{0,A}(u,\xh) $ is
\eq{zz6}{\p_u  a^+_{0A}(u,\xh)  \sim   (u-u_0-i\eps)^{-{d-2\over 2} }  +\ldots }
where $\ldots$ are terms that die off more rapidly at large $u$.  This falloff is needed for the soft theorem and can be seen in explicit examples of the type considered in $d=4$.  So, for example, it follows that
$\int\! du \nabh^A \p_u a^+_{0A}(u,\xh)  \delta \Phi(\xh) =0$ for $d>4$ simply by closing the contour.    As we show momentarily, a nonzero contribution will arise from a subleading term, according to $\Phi_n \sim   u^n$.

\subsection{Goldstone mode}

The Goldstone mode is taken to obey $\nabla^2 \Phi$ near the boundary; the justification for this will become clear once we study the theory at spatial infinity, where $\nabla^2 \Phi=0$ is required for invariance under large gauge transformations.\footnote{In fact, we only require $\nabla^2 \delta \Phi=0$ where $\delta \Phi $ is obtained from a large gauge transformation, but we don't lose any information by considering the stronger condition $\nabla^2 \Phi=0$.}   Given the expansion \rf{zz2}, we wish to obtain equations relating $\Phi_{n>0}$ to $\Phi_0(\xh)$.     To avoid the complication of non-commuting covariant derivatives, it's convenient to use coordinates for which the metric on $\Ic^+$ involves a flat celestial plane rather than a celestial sphere.\footnote{The distinction is analogous to Poincar\'e versus global coordinate for AdS.}   It will be straightforward to reconstruct  the desired expressions in the usual sphere coordinates.  

We first collect some needed formulas for celestial plane coordinates.  Writing the Minkowski metric as $ds^2 = -(dX^0)^2 + \sum_{i=1}^{d-1} (dX^i)^2$ we write 
\eq{zp1}{ X^\mu = {r\over 2} \left( 1+ x\cdot x+{u\over r} ,2x^A,1-x\cdot x-{u\over r}   \right) }
with $x^A\in \mathbb{R}^{d-2}$ with metric $\delta_{AB}$. The metric in these coordinates is 
\eq{zz7}{ ds^2 =-dudr +r^2 dx^A dx^A}
We write the Cartesian components of a  null momentum  as
\eq{zp2}{ q^\mu =\omega_q \hat{q}_\mu =  \omega_q \left( {1+ y\cdot y\over 2},y^A,{1-y\cdot y\over 2}\right)   }
and the $d-2$ transverse polarization vectors have Cartesian components 
\eq{zp3}{ \veps^A_\mu = \p^A \hat{q}_\mu = (-y^A,\delta^A_B,-y^A)~.  }
Note that the polarization index, previously denoted by the helicity $\alpha$, is here denoted by $A$, corresponding to a basis of linear polarizations. 
There are analogous expressions with $u$ replaced by $v$. 

 The wave equation for $\Phi$ in these coordinates is
\eq{zz8}{\nabla^2 \Phi = -2 r^2\p_r \p_u \Phi  -(d-2)r\p_u \Phi+\nabh^2 \Phi=0 }
with $\nabh^2$ the  Laplace operator on $ \mathbb{R}^{d-2}$. 
Inserting the  expansion  \rf{zz2}  implies 
\eq{zz9}{ \p_u \Phi_n =- {1\over 2n-d+2} \nabh^2 \Phi_{n-1}}
which is solved as 
\eq{zz10}{  \p^n_u \Phi_n =   {1 \over 2^n} {\Gamma\left( {d-2\over 2}-n\right)  \over \Gamma\left( {d-2\over 2}\right)   } (\nabh^2)^n \Phi_0   }
We note in particular that the $\Phi_n$ term in \rf{zz2} thus contains a dependence ${u^n \over r^n}$, along with lower $u$ powers. Given the form of $\p_u a^+_{0A}$ in \rf{zz6} we see that the $n={d-4\over 2}$ term in the $\Phi$ expansion is just right to give a nonzero $u$ integral along with a finite large $r$ limit.  We in particular have 
\eq{zz11}{  \Phi_{d-4\over 2} = {1\over 2^{{d-4\over 2}} \Gamma^2\left( {d-2\over 2}\right)} u^{{d-4\over 2}}  (\nabh^2)^{{d-4\over 2}}  \Phi_0 + \ldots  }
where $\ldots$ denote lower powers of $u$. This will give the following contribution to the action variation
\eq{zz12}{ \delta I = \int_{\Ic^+} d^3x  \nabh^A \p_u a^+_{0A} \delta \Phi_{d-4\over 2} + (\Ic^-) }
In principle there are other contributions coming from combining subleading in $r$ contributions to $a^+_A$ and $\Phi$ such as to give a nonzero $u$ integral.  In fact, we will find that \rf{zz12} gives the full contribution needed to produce the soft theorem, but we do not have a direct argument as to why the other potential contributions are absent; we leave this for future work. 

\subsection{Ward identity}

As we will discuss in  section \rff{Sec: Variations at spatial infinity}, careful analysis including potential contributions at spatial infinity, establishes that the partition function is invariant under large gauge transformations.  The decomposition
\eq{zz13}{
    \overline A{_{0A}^\pm}(u, \hat x) = \hatline{A}{^\pm_{0A}}(u, \hat x) + C^\pm_A(\hat x)
}
where $C^\pm_A= \nabla_A \Phi_\pm$ (to lowest order in $\Phi$) implies that a LGT shifts the Goldstone mode $\Phi$ along with acting on the gluons and other charged matter.  By the same logic as in $d=4$, the decomposition \rf{zz13} may be used to relate the change in the partition function under a shift of  $\Phi$ to the change under a shift of $ \hatline{A}{^\pm_{0A}}(u, \hat x) $.   Altogether, this yields a Ward identity for the partition function involving only the gluons and matter fields, with $\Phi$ set to zero.  The one technical complication that arises for $d>4$ is that both leading $\lambda_0$ and subleading $\lambda_{{d-4\over 2}}$ components of the gauge parameter act on the fields, and are related as in \rf{zz11}.   Here we are taking the gauge parameter to obey $\nabla^2 \lambda=0$ in a vicinity of the boundary; this condition is required at spatial infinity in order to avoid unwanted boundary terms in the gauge variation of the action (see section \rff{Sec: Variations at spatial infinity}), and extends by continuity to a vicinity of null infinity.  

We thus have the Ward identity \rf{ST3} where the charge operators are now
\eq{zz14a}{
    i\hat Q^a_+(\hat x) &= \phantom{-}\int_{\scrI^+}\dr^{d-1} y \Bigg[ -\frac{1}{2}\delta^a_c {u^{{d-4\over 2}}  (\nabh_y^2)^{{d-4\over 2}} \over 2^{{d-4\over 2}} \Gamma^2\left( {d-2\over 2}\right)} \hat\nabla_{yA}\delta^{(d-2)}(\hat y - \hat x) \cr
    & \qquad\quad\quad\quad\quad\quad\quad\quad\quad + g\struct{a}{b}{c}\hatline{A}\subsuper{0A}{b-}(u, \hat y)\delta^{(d-2)}(\hat y - \hat x) \Bigg] \frac{\delta}{\delta \hatline{A}\subsuper{0A}{c-}(u, \hat y)},\cr
    i\hat Q^a_-(\hat x) &= -\int_{\scrI^-}\dr^{d-1} y\Bigg[ -\frac{1}{2}\delta^a_c {v^{{d-4\over 2}}  (\nabh_y^2)^{{d-4\over 2}} \over 2^{{d-4\over 2}} \Gamma^2\left( {d-2\over 2}\right)} \hat\nabla_{yA}\delta^{(d-2)}(\hat y + \hat x) \cr
    & \qquad\quad\quad\quad\quad\quad\quad\quad\quad  + g\struct{a}{b}{c}\hatline{A}\subsuper{0A}{b+}(v, \hat y) \delta^{(d-2)}(\hat y + \hat x) \Bigg] \frac{\delta}{\delta \hatline{A}\subsuper{0A}{c+}(v, \hat y)}.
}
and we can again use that the two charges  are equal.  
The same steps leading to \rf{SG4} now yield the Ward identity 
\eq{zz14b}{
    -&\frac{1}{2} { (\nabh_y^2)^{{d-4\over 2}} \over 2^{{d-4\over 2}} \Gamma^2\left( {d-2\over 2}\right)} \hat\nabla_{yA}\Big[\int_{\scrI^+}\dr^{d-1} y' \delta^{(d-2)}(\hat y - \hat y')u'^{{d-4\over 2}}  W_\text{out}^{aA;b_1B_1;\ldots}(y', x_1,\ldots) \cr
    & \quad \quad \quad \quad \quad \quad \quad~~ + \int_{\scrI^-}\dr^{d-1} y'\delta^{(d-2)}(\hat y + \hat y')  v'^{{d-4\over 2}}W_\text{in}^{aA;b_1B_1;\ldots}(y', x_1,\ldots) \Big]\cr
    &= g \sum_{n=1} \struct{a}{b_n}{c} \delta^{(d-2)}(\hat y - \hat\eta_n\hat x_n) W^{b_1B_1;\ldots;cB_n;\ldots}(x_1,\ldots)
}

\subsection{Solution of Ward identity}

We use that the Green's function on $\mathbb{R}^{d-2}$ obeying
\eq{zz14c}{ \nabh_y^A G_A(y)  =  \delta^{d-2}(y-y_i)}
is given by 
\eq{zz14d}{ G_A(y)& = -{ \Gamma\left({d-4\over 2}\right) \over 4 \pi^{{d-2\over 2}} } \nabh_{yA} {1\over |y-y_i|^{d-4} }  }
Taking into account the Laplace operators appearing in \rf{zz14b} we will
 need that  the solution of
\eq{zz14e}{ (\nabh_y^2)^{{d-4\over 2}} \nabh_y^A F_A(y)=\delta^{d-2}(y-y_i) }
is
\eq{zz14f}{ F_A(y) = {(-1)^{d/2}\over (4\pi)^{{d-2\over 2}} \Gamma\left({d-2\over 2}\right)}   \nabh_A  \ln|y-y_i|^2  }
Next, using that  the points $(y, y_i)$  on $\mathbb{R}^{d-2}$ corresponds to null momenta $(q,p_i)$, and using the expressions \rf{zp2}-\rf{zp3} we have
\eq{zz14g}{   \nabh_A \ln|y-y_i|^2 =\veps^{*B}_A{p_i \cdot \veps^B(\qh)\over p_i\cdot \hat{q}} }
and thereby arrive at 
\eq{zz14h}{  &\int_{\Ic^+} du  u^{{d-4\over 2}}  W_\text{out}^{aA;b_1B_1;\ldots}(y, x_1,\ldots) \cr
& \quad  \quad   \quad =-{g\over 4\pi} { (-1)^{d/2} \Gamma\left({d-2\over 2}\right)\over (2\pi)^{{d-4\over 2}}  }\veps^{*B}_A \left( \sum_{n=1} f^a_{~b_n c}  {p_n \cdot \veps^B(\qh)\over p_n\cdot \hat{q}} \right) W^{b_1B_1;\ldots;cB_n;\ldots}(x_1,\ldots)  }
As was already discussed in $d=4$, the integral on the LHS  is only conditionally convergent, and is defined as $\int\!_{\Ic^+} du(\ldots)  = \lim_{L\rt \infty} \int_{-L}^L\! du (\ldots)  $.  With this understanding we have  (see appendix \ref{Sec: Conditionally convergent integral})
\eq{zz14i}{   \lim_{\omega \rt 0^+}\int_{\Ic^+} du  e^{i\omega u}  u^{{d-4\over 2}}  W_\text{out}^{aA;b_1B_1;\ldots}(y, x_1,\ldots) =2 \int_{\Ic^+} du  u^{{d-4\over 2}}  W_\text{out}^{aA;b_1B_1;\ldots}(y, x_1,\ldots) }
The integral on the LHS is what arises from the Ward identity, while on the RHS we have the soft limit of the amplitude.  

The Ward identity relation \rf{zz14h} was obtained by working with the celestial plane rather than the celestial sphere (this avoids having to deal with non-commuting covariant derivatives in expressions like \rf{zz14d} ).  But \rf{zz14h} is valid as stated\footnote{except that we will relabel the polarization vector as $\veps^A_\mu \rt \veps^\alpha_\mu$ and the polarization vector $\eps^{*B}_A $ will acquire a hat, as defined in \rf{zz20}. } on the sphere as well, and so in the remainder of this section we revert back to the sphere.

In $d$-dimensions, the momentum space pole in the amplitude corresponds to a $(u-i\eps)^{-{d-2\over 2}}$ large $u$-falloff on $\Ic^+$. Suppressing dependence on the non-soft gluons we can then write
\eq{zz14aa}{ W_{{\rm out}}^{a,A}(y) = { w_{{\rm out}}^{a,A}(\yh) \over (u-i\eps)^{{d-2\over 2}}  }+ \ldots  }
Using that terms with faster falloff contribute nothing to the integral we have 
\eq{zz15}{  \lim_{L\rt \infty} \int_{-L}^L \! du  u^{{d-4\over 2}} W_{{\rm out}}^{a,A}(y)  = i\pi w_{{\rm out}}^{a,A}(\yh)  }
and we then use \rf{zz14h} to obtain the Ward identity in terms of $w_{{\rm out}}^{a,A}(\yh) $.
Now we need to work out an expression for the soft part of the S-matrix element.  One could proceed as in section \ref{Sec: Soft gluon theorem}, but it is instructive to see how one can instead use that the S-matrix is equal to the (normal ordered) partition function, where the mode coefficients are promoted to operators. The connected part is given by  $\hat{S}= : \ln \hat{Z}: $.   Again suppressing dependence on the non-soft gluons, we need\footnote{ In this section we suppress the positive/negative frequency labels to reduce clutter.}
\eq{zz16}{ \langle 0| \hat{a}^a_\alpha(\qv) i\int_{\Ic^+} \! du d^{d-2}\yh W_{{\rm out}}^{b,A}(y) \hat{\Ab}{_{0A}^b}(u,\yh)|0\rangle  }
corresponding to the emission of a gluon with polarization $\alpha$ and color index $a$, where the Fourier coefficients in $\Ab_{0A}^a$ have been promoted to operators obeying 
\eq{zz17}{ [\hat{a}_\alpha^a(\qv),\hat{a}_\beta^{b}(\qv')]  &=(2 \pi)^{d-1} 2 \omega_{\vec{q}} \delta^{(d-1)}\left(\vec{q}-\vec{q}^{\prime}\right)  \cr
& =   (2 \pi)^{d-1} 2  \omega^{3-d}  \delta(\omega-\omega') \delta^{(d-2)}(\qh-\qh')    }
The mode expansion of the free gluon field is 
\eq{zz18}{ A^a_\nu(x)= \sum_{\alpha} \int \frac{d^{d-1} q}{(2 \pi)^{d-1}} \frac{1}{2 \omega}\left[\varepsilon_\nu^{* \alpha}(\vec{q}) a^a_\alpha(\vec{q}) e^{i q \cdot x}+\varepsilon_\nu^\alpha(\vec{q}) {a^a_\alpha}^\dagger(\vec{q})e^{-i q \cdot x}\right] }
where $\alpha$ runs over the $d-2$ transverse polarizations.  Near $\Ic^+$ the angular components have leading falloff $A_A \approx A_{0A} r^{-{d-4\over 2}}$ with 
\eq{zz19}{    A^a_{0A}(x) \approx {(-2\pi i)^{{d-2 \over 2}} \over 2 (2\pi)^{d-1}  }  \int_0^\infty \! d\omega \omega^{{d-4\over 2}} \big[ \vepsh^{*\alpha}_A( \omega \xh) e^{-i\omega u} a^a_\alpha( \omega \xh) - (-1)^{{d-4\over 2}}\vepsh^\alpha_A( \omega \xh) e^{i\omega u} {a^a_\alpha}^\dagger(\omega \xh)    \big]  }
as is obtained via saddle point evaluation of the $\omega $  integral and where 
\eq{zz20}{  \vepsh^\alpha_A= {1\over r} {\p x^\nu \over \p \xh^A} \veps^\alpha_\nu }
We can now evaluate \rf{zz16} as $\omega \rt 0$ as
\eq{zz21}{ \langle 0| \hat{a}^a_\alpha(\qv) i\int_{\Ic^+} \! du d^{d-2}\yh W_{{\rm out}}^{b,A}(y) \hatline{A}{_{0A}^b}(u,\yh)|0\rangle  & = -  {(-1)^{d/2}  (2\pi)^{d/2} \over \Gamma\left({d-2\over 2}\right) } {1\over \omega_q}  \vepsh^\alpha_A(\hat{q}) w_{\rm out}^{a,A}(\qv) +O(\omega_q^0)   }
 where we used  \rf{zz14i} and \rf{zz15}.     Putting these formulas together and using $\vepsh^{*\alpha}_A \vepsh^{\beta A} =\delta_{\alpha\beta} $ we arrive at
 \eq{zz22}{
    \lim_{\omega\arrow 0^+}\omega \mathcal{A}_\text{out}^{\alpha;b_1B_1;\ldots}(\omega \hat q; x_1,\ldots) = -ig\sum_{k=1}\struct{a}{b_k}{c}\frac{\epsilon^\alpha(\hat y)\cdot n(\hat\eta_k\hat x_k)}{n(\hat y)\cdot n(\hat\eta_k \hat x_k)}W^{b_1B_1; \ldots; cB_k;\ldots}(x_1, \ldots)
}
which agrees with  \eqref{kk4}.

\subsection{Variations at spatial infinity }
\label{Sec: Variations at spatial infinity}

To this point, we have not considered the role played by spatial infinity in our analysis. Previous work in Abelian gauge theory has noted that the behavior of the fields at spatial infinity can be closely related to the question of antipodal matching which was crucial in section \ref{Sec: Soft gluon theorem} to obtaining a useful Ward identity from large gauge invariance \cite{Campiglia:2017mua,Esmaeili:2019hom}. However, these results generally rely on the particular choice of Lorenz gauge, at least asymptotically, which has the key feature that residual gauge transformations obey the free, massless wave equation. Solutions of the wave equation going as $r^0$ near null infinity in $d = 4$ always obey the antipodal matching condition.\footnote{It has been shown that it is sufficient for a field to only obey the wave equation in the vicinity of spatial infinity to find antipodal matching \cite{Campiglia:2017mua}.} This approach does not generalize well to non-Abelian gauge theory because neither $\p_\mu A^\mu = 0$ nor $D_\mu A^\mu = 0$ gauge imply that the residual gauge transformations obey the free wave equation.

Nonetheless, as we will now show, the antipodal matching condition can be understood directly from the action \eqref{YM3}   extended to  dimensions $d > 4$. In our presentation, the antipodal matching condition will arise by demanding that the variation of the Goldstone mode receives no contribution from spatial infinity.  Such a contribution, if present, would spoil the key relation between variations of the action with respect to the Goldstone mode and the dynamical data $\hatline{A}\subsuper{0A}{a}$.

We resolve spatial infinity by working in the dS slicing of the Rindler wedge in which the coordinates $(\rho, \tau, \hat x)$ are related to polar coordinates in Minkowski space by
\eq{spat1}{
    \tau = \frac{t}{\sqrt{r^2 - t^2}}, \ \ \ \rho = \sqrt{r^2 - t^2}.
}
This chart covers the region $r > |t|$ and the metric is given by
\eq{spat2}{
    \dr s^2 &= \dr \rho^2 + \rho^2 h_{\alpha\beta}\dr x^\alpha \dr x^\beta\cr
    h_{\alpha\beta}\dr x^\alpha \dr x^\beta &= \frac{\dr\tau^2}{1 + \tau^2} + (1 + \tau^2)\dr\Omega_{d-2}^2
}
so $h_{\alpha\beta}$ is the dS$_{d-1}$ metric. Spatial infinity is reached by taking $\rho\arrow\infty$.
In these coordinates, we follow the  analysis of \cite{Esmaeili:2019hom}. The field strength components are assumed to have the following large $\rho$ falloffs
\eq{zx1}{ F_{\rho \alpha} \sim F_{\alpha\beta} \sim \rho^{3-d}~. }
The $F_{\rho \alpha} $ falloff corresponds to that produced by a freely propagating massive charged particle.  The $F_{\alpha \beta}$ falloff is restrictive enough to disallow the presence of magnetically charged objects.  A simplification of considering $d>4$ is that the falloffs imply that the action effectively becomes quadratic near spatial infinity, in the sense that cubic and quartic terms in the action falloff too rapidly to contribute to boundary terms as $\rho \rt \infty$.  

We will also need the falloffs of the gauge potential, which we write as 
\eq{zx2}{ A_\rho(\rho,x^\alpha) & = \rho^{3-d} A_\rho^{(d-3)}(x^\alpha) + \nabla_\rho \Phi(\rho,x^\alpha) + \ldots \cr
 A_\alpha(\rho,x^\alpha) & = \rho^{4-d} A_\alpha^{(d-4)}(x^\alpha) +  \rho^{3-d} A_\alpha^{(d-3)}(x^\alpha)  + \nabla_\alpha \Phi(\rho,x^\alpha) + \ldots }
 The falloffs \rf{zx1} require that $A_\alpha^{(d-4)}(x^\alpha)$ be pure gauge on dS$_{d-1}$, so we write 
 \eq{zx3}{A_\alpha^{(d-4)}(x^\alpha) =\nabla_\alpha \Lambda(x^\alpha)~.  }
We include this since it is typically present, for example in the standard  Coulomb solution for a boosted point charge.  The Goldstone mode $\Phi$ has leading $r^0$ behavior at null infinity, and to smoothly match that behavior it should have leading $\rho^0$ behavior at spatial infinity.  We assume a series expansion near spatial infinity,
\eq{zx4}{ \Phi(\rho,x^\alpha) = \Phi_0(x^\alpha) + \rho^{-1} \Phi_1(x^\alpha)+ \ldots}

Having specified the falloffs we consider boundary terms arising from the on-shell variation of the action, where the boundary is placed at fixed $\rho$.  It is sufficient for our purposes to consider the variation around a background with $\Phi=0$.  The variation of the Yang-Mills terms yields
\eq{zx5}{\delta I_{\rm YM}=  \delta\left[ -{1\over 4}  \int\! d^d x\sqrt{g} \tr F^{\mu\nu}F_{\mu\nu}   \right]& = - \int d^{d-1}x \sqrt{h} \rho^{d-3} h^{\alpha\beta} \tr  F_{\rho \alpha} \delta A_\beta~.   } 
For $d>4$ the falloff conditions imply that the contributions come only from the part of $F_{\rho \alpha}$ linear in the gauge field and the Goldstone part of $\delta A_\beta$. It's also clear that the gauge fixing and ghost terms don't contribute, as they involve the product of at least two non-Goldstone fields.   We therefore find the following variation 
\eq{zx6}{ \lim_{\rho \rt \infty} \delta I  =-\int \!d^{d-1}x \sqrt{h} h^{\alpha \beta} \big( (4-d)\nabla_\alpha \Lambda - \nabla_\alpha A_\rho^{(d-3)} \big) \nabla_\alpha \delta  \Phi_0 }
Integrating by parts on dS$_{d-1}$\footnote{ One can check that the resulting boundary terms vanish under our falloff assumptions.} we find a nonzero term proportional to $\nabla^2_{\rm dS} \delta  \Phi_0$.    For the reasons explained above, such a term will modify the derivation of the Ward identity unless we impose the condition
\eq{zx7}{\nabla^2_{\rm dS}  \Phi_0 =0~.}
As discussed in \cite{Campiglia:2017mua,Esmaeili:2019hom} admissible solutions of this equation obey an antipodal matching relating their early and late time behavior,
\eq{zx8}{  \lim_{\tau=\infty} \Phi_0(\tau,\xh) = \lim_{\tau=-\infty} \Phi_0(\tau,-\xh)~.  }
This in turn implies a corresponding antipodal match between $\Phi$ on $\Ic^+$ and $\Ic^-$
\eq{zx9}{ \Phi_0(\xh) \big|_{\Ic^+} = \Phi_0(-\xh) \big|_{\Ic^-}} 
as follows from the fact that the dS slices approach $\Ic^\pm$  as $\tau \rt \pm \infty$.
To summarize, assuming we impose \rf{zx7}, which imposes the antipodal match on $\Phi$ at $\Ic^\pm$, we avoid an unwanted boundary variation at spatial infinity and therefore  an unwanted contribution to the  Ward identity.

\section{Future directions}
\label{Sec: Future directions}

In this work we have described the path integral dictionary for the case of bulk non-Abelian gauge fields. This path integral provides a transparent framework for addressing questions in flat space holography, some of which we have addressed here. In this section we offer some comments on other open problems which might be usefully addressed within the path integral framework.

\vspace{.2cm}
\noindent 
{{\bf{\large{\underline{Massive fields}}}}}
\vspace{.2cm}

Throughout this work we have focused exclusively on massless fields, whose asymptotic data is supported on null infinity. However, a genuine Carrollian dual should be able to incorporate the effects of massive bulk fields. The asymptotic data can be represented on a hyperbolic resolution of timelike infinity, and we give some preliminary comments on this story in appendix \ref{Sec:Massive particles}. However, this construction does not result in a unified description of massive and massless fields since the asymptotic data are not supported on the same boundary manifold. The precise junction conditions fusing null infinity to the hyperboloid at timelike infinity are also not completely clear, and seemingly sensitive to the IR regulator of the theory since the hyperbolic coordinates cover a portion of null infinity as well.

\vspace{.2cm}
\noindent 
{{\bf{\large{\underline{IR regulation and loop corrections}}}}}
\vspace{.2cm}

Throughout this work we have  in mind working in dimensions $d > 4$ where IR issues do not appear. However, a direct IR regulation in $d = 4$ should be possible, and would be expected to introduce new effects. For instance, amplitudes computed by expanding around $d = 4$ are known to obey loop-corrected soft theorems. This implies that IR regulating the Carrollian partition function directly in $d = 4$ must modify the argument described in sections \ref{Sec: Ward identity of large gauge transformations} to \ref{Sec:Charge algebra}.

This may occur due to additional counter terms which explicitly break the large gauge symmetry, e.g. below \eqref{YM6} we observed the need for a $\frac{1}{2}(a\subsuper{a\ln}{0A})^2$ term to cancel a log divergences due to the $\ln |u|$ behavior of the fluctuating field at large $u$. In the presence of loop corrections one expects, in dimensional regularization, momentum space behavior like $\frac{1}{\omega^{1+\epsilon}}$, which expands to terms of the form $\frac{1}{\omega}(\ln\omega)^n$. These Fourier transform via \eqref{r19} to boundary correlators with behavior $(\ln u)^n$, leading to additional divergent terms in the action variation \eqref{YM6} that need to be canceled. It is possible that these additional counter terms break the large gauge symmetry.

Alternatively, it may be the case that the calculation carried out for $d > 4$ in section \ref{Sec: Variations at spatial infinity} is significantly modified when working directly in $d = 4$. There, operators supported at spatial infinity may enter the story and deform the relation \eqref{YM11}. This would lead directly to a deformation of the charges \eqref{SG3}, and propagate to a deformation of the soft theorem.

\vspace{.2cm}
\noindent 
{{\bf{\large{\underline{Equations of motion and subleading field components}}}}}
\vspace{.2cm}

In section \ref{Sec: Soft gluon theorem} we restricted attention to the leading soft gluon theorem, which is directly tied to invariance under the large gauge transformations considered herein. Subleading soft theorems, for photons and gluons, have been discussed in many works, including \cite{Low:1958sn,Bern:2014vva,Lysov:2014csa,Campiglia:2016hvg,Mao:2017tey,Himwich:2019dug,AtulBhatkar:2019vcb,Campiglia:2021oqz,Peraza:2023ivy,Strominger:2021mtt,Choi:2024ygx,Choi:2024ajz,Choi:2024mac}. However, the precise connection to large gauge transformations is not yet fully established.\footnote{See \cite{Kraus:2024gso} for a discussion in the current framework.} Some of these recent approaches involve subleading terms in the $1/r$ expansion of the fields, schematically
\eq{FD1}{
    \phi = \frac{1}{r}\phi_1 + \frac{1}{r^2}\phi_2 + \cdots.
}
Here we have only discussed the leading term, which holds our fixed data as in \eqref{r2}. The subleading terms in \eqref{FD1} are not immediately accessible in the present framework since we can only study correlators of operators dual to variations of our data, i.e. the unfixed frequency component of $\phi_1$.

However, since the equations of motion hold within the path integral, it can be used to give a definition for the subleading operators. Expanding the equations of motion about null infinity generally produces iterative equations of the form
\eq{FD2}{   
    \p_u\phi_n(u, \hat x) = F[\phi_1,\ldots\phi_{n-1}](u, \hat x)
}
where $F$ is some local polynomial in the higher coefficients. The relation \eqref{FD2} holds as an operator equation inside the path integral, and recursive application allows one to define the expectation of $\phi_n$ in terms of some polynomial constructed from $\phi_1$, which we have direct access to from variations of the Carrollian partition function.

However, applying this idea is subtle. Firstly, it relies on the field obeying a specified expansion at large $r$ which may be corrected, particularly at loop order. Secondly, \eqref{FD2} is explicitly a product of operators at a coincident point, and hence the operator on the right must be defined carefully in some manner. Certainly, one cannot simply assume that $\langle F[\phi_1]\rangle = F[\langle\phi_1\rangle]$.

\section*{Acknowledgments}

We thank Eric D'Hoker for discussions.  P.K. is supported in part by the National Science Foundation grant PHY-2209700.

\appendix

\section{Conventions}
\label{Sec: Conventions}

The Lie algebra is written 
\eq{q1}{ [t_a,t_b]=i f^c_{~ab} t_c}
with a field  $\phi^\mathcal{R}$ in representation $\mathcal{R}$ transforming as 
\eq{q2}{ \phi'^{\mathcal{R}}  = U^\mathcal{R} \phi^R~,\quad U^\mathcal{R} = e^{ig \lambda^\mathcal{R}}~,\quad \lambda^\mathcal{R} = \lambda^a t_a^\mathcal{R} }
where $g$ denotes the Yang-Mills coupling.  
 The covariant derivative is
\eq{con3}{
    D_\mu \phi^\mathcal{R} = \p_\mu \phi^\mathcal{R} - igA_\mu^\mathcal{R}\phi^\mathcal{R}
}
where the gauge connection is $A_\mu^\mathcal{R} = A_\mu^a t_a^\mathcal{R}$. We will often omit the superscript when there should be no confusion about the representation. The transformation of the connection is
\eq{con4}{
    A_\mu' = U^{-1}(A_\mu + \p_\mu)U = D_\mu\lambda + O(\lambda^2) = (\p_\mu\lambda^a + g\struct{a}{b}{c}A^b_\mu \lambda^c)t_a + O(\lambda^2).
}

The field strength is
\eq{con5}{
    \relax[D_\mu, D_\nu] &= -ig F_{\mu\nu}\\
    F_{\mu\nu} &= \p_\mu A_\nu - \p_\nu A_\mu - ig[A_\mu, A_\nu]
}
with $F_{\mu\nu}= F_{\mu\nu}^a t_a$.   We also use the notation
\eq{q5}{ \tr (F_{\mu\nu} F^{\mu\nu} ) =F^a_{\mu\nu} F^{a\mu\nu}~. }
The generators in the adjoint representation are $(t_b^\text{adj})^a_c = i\struct{a}{b}{c}$.

\section{LSZ = Carrollian partition function, the variational principle, and background field gauge}
\label{Sec: LSZ = AFS}

Given a list of Feynman rules for computing correlators, one has an unambiguous prescription, via LSZ reduction, for computing S-matrix elements. These rules  depend only on the bulk terms in the action, but the definition of the Carrollian partition function requires a specification of boundary terms. So one may question which precise boundary terms are guaranteed to reproduce the S-matrix elements computed by the Feynman rules. In \cite{Kim:2023qbl} a proof, valid to all orders in perturbation theory, was given that the boundary terms described in section \ref{Sec: Review} are the correct ones to make this correspondence. In this appendix, we review the argument in a way that emphasizes the role played by a good variational principle.\footnote{The argument here can also be formulated in AdS where it would present a view of  holographic renormalization terms similar to that advocated in \cite{Papadimitriou:2007sj}.} For this purpose, it will be sufficient to suppose we have a collection of fields $\phi$ whose type will not matter in the following.

The argument is explicitly perturbative about free field theory, though it will be all-orders, and so we suppose the action may be written
\eq{AFS1}{
    I[\phi, \overline\phi, J] = I_0[\phi, J] + I_\text{int}[\phi] + I_\text{bndy}[\phi, \overline\phi]
}
where $I_0$ is the quadratic action of a free field together with an added $\int J\phi$ source term and with boundary terms  described in section \ref{Sec: Review}. This is equivalent to choosing $I_0$ to be the free action supplying a good variational principle with respect to the scattering boundary conditions described there. The background field $\overline\phi$ is the free field obeying the boundary conditions $\overline\phi_1^\pm$ described in section \ref{Sec: Review}. We assume that the variation of $I_\text{int}$ contributes no boundary terms to the on-shell value of $\delta I$. In operator language, this would mean we assume the canonical commutation relations are those implied by the free action $I_0$ together with the boundary terms $I_\text{bndy}$. For \eqref{AFS1} to furnish a good variational principle, we would need to choose $I_\text{bndy}$ to depend only on $\overline\phi$, but will not yet impose this condition.

We define the generating functional
\eq{AFS2}{
    Z[\overline\phi, J] = \int_{\overline\phi}\D\phi e^{iI[\phi, \overline\phi, J]}.
}
This object serves to interpolate between the special case $Z[0, J]$, which is familiar as the generating function of bulk $n$-point Green's functions from which we extract S-matrix elements via the LSZ reduction\footnote{One might worry that the residual boundary condition setting the positive (negative) frequency component of the fields to zero in the far past (future) is non-standard. However, these are the precise boundary conditions obeyed by the Feynman Green's functions. One may also note that in Euclidean signature, these boundary conditions translate to the familiar requirement that the fields fall off at large Euclidean distance. This latter perspective was considered in \cite{Jain:2023fxc}.}, and the case $Z[\overline\phi, 0]$, which would be the Carrollian partition function if the boundary term were chosen such that $I$ furnished a good variational principle.

In the case $Z[0, J]$  we have, by definition,
\eq{AFS3}{
    Z[0, J] &= e^{\frac{i}{2}\int JG_FJ}\sum_{n=3}\frac{1}{n!}\int G^\text{amp}_n\left( i\int G_FJ \right)^n
}
where $G_F$ is the free field Feynman 2-point Green's function and $G^\text{amp}_n$ is the $n$-point amputated correlator. In the case $Z[\overline\phi, 0]$, we have
\eq{AFS3a}{
    Z[\overline \phi, 0] = e^{iI_0[\overline\phi, 0]}\sum_{n=3}\frac{1}{n!}\int W_n \overline\phi\subsuper{}{n}.
}
We view \eqref{AFS3} and \eqref{AFS3a} as the definition of what we mean by $G^\text{amp}_n$ and $W_n$ and our task will be to find the relation between them. In the case of \eqref{AFS3a}, we note that $iI_0[\overline\phi, 0] \sim \int\dr^3 p b^\dag b$, so the exponential factor is nothing but the generating function for the free S-matrix, $\hat S = \hat 1$, and thus supplies the free component of $\hat S = \hat 1 + i\hat T$.

In \eqref{AFS2} it will be useful to shift the boundary conditions away by writing
\eq{AFS4}{
    \phi &= \overline\phi_J + \tilde \phi,\cr
    \overline\phi_J &\equiv \overline\phi + i\int G_FJ
}
so that $\overline\phi_J$ is the unique sourced free field obeying the boundary conditions. With this shift, we note that, since it is a quadratic form, the free action obeys
\eq{AFS5}{
    I_0[\overline\phi_J + \tilde\phi, J] = I_0[\overline\phi, 0] + \frac{1}{2}\int JG_FJ + I_0[\tilde\phi, 0].
}
No boundary terms coupling $\overline\phi$ to the fluctuations $\tilde\phi$ arise in this expansion specifically because $I_0$ was chosen to be the free action supplying a good variational principle.

Using the shift \eqref{AFS4} and the expansion \eqref{AFS5} in the path integral \eqref{AFS2}, we find
\eq{AFS6}{
    Z[\overline\phi, J] = e^{iI_0[\overline\phi, 0] + \frac{i}{2}\int JG_FJ}\int_0\D\tilde\phi \exp\left( iI_0[\tilde\phi, 0] + iI_\text{int}[\overline\phi_J + \tilde\phi] + iI_\text{bndy}[\overline\phi_J + \tilde\phi, \overline\phi] \right).
}
Now, were it the case that the integrand of \eqref{AFS6} depended only upon the combination $\overline\phi_J$, it would necessitate that
\eq{AFS7}{
    \sum_{n=3}\frac{1}{n!}\int G^\text{amp}_n \overline\phi\subsuper{}{n} = \sum_{n=3}\frac{1}{n!}\int W_n \overline\phi\subsuper{}{n}.
}
In other words, \eqref{AFS3} must equal \eqref{AFS3a} under the replacement $i\int G_FJ \arrow \overline\phi$. The precise relation between $G^\text{amp}_n$ and $W_n$ is then obtained by the calculation described for the scalar in section \ref{Sec: Review} and for the gauge field in section \ref{Sec: Dual operators and boundary correlators}.

The term spoiling this conclusion is $I_\text{bndy}$, which is allowed to depend separately on $\overline\phi$, breaking the symmetry between $\overline\phi$ and $i\int G_F J$. One option would be to suppose that $I_\text{bndy} = I_\text{bndy}[\overline\phi_J + \tilde\phi]$, in which case the argument would proceed as planned, though this would also imply that $Z[0, J]$ is computed via Feynman rules with non-trivial vertices supported on the boundary of spacetime, which is non-standard and undesirable. We therefore require $I_{\rm bndy}= I_{\rm bndy}[\overline\phi]$, corresponding to a good variational principle.

Hence we find that the path integral with boundary conditions will compute the same generating functional determined by the usual Feynman rules only when we choose the boundary terms to be those that produce a good variational principle. However, demanding a good variational principle alone does not completely determine the boundary terms since it allows for an arbitrary $I_\text{bndy} = I_\text{bndy}[\overline\phi]$.  Demanding that such a term be local on the boundary, it will contribute only contact terms in correlators, so has no effect on non-collinear  S-matrix elements.

To give a precise account of the effects of including a nonzero $I_\text{bndy}[\overline\phi]$, define the modified generating function coefficients $W'_n$ by
\eq{AFS8}{
    Z[\overline\phi, 0] = e^{iI_0[\overline\phi, 0] + iI_\text{bndy}[\overline\phi]}\sum_{n=3}\frac{1}{n!}\int W'_n\overline\phi\subsuper{}{n}.
}
Examining \eqref{AFS6} informs us that the correct correspondence to standard amputated amplitudes would be
\eq{AFS9}{
    \sum_{n=3}\frac{1}{n!}\int G^\text{amp}_n\overline\phi\subsuper{}{n} = \sum_{n=3}\frac{1}{n!}\int W'_n\overline\phi\subsuper{}{n}
}
since it's the part of $Z[\overline\phi, J]$ depending only on the combination $\overline\phi_J$ which is required to match in the two limits. In terms of the original coefficients $W_n$, this is
\eq{AFS10}{
    \sum_{n=3}\frac{1}{n!}\int G^\text{amp}_n\overline\phi\subsuper{}{n} = e^{-I_\text{bndy}[\overline\phi]}\sum_{n=3}\frac{1}{n!}\int W_n\overline\phi\subsuper{}{n}.
}
Throughout this work, we choose $I_\text{bndy}[\overline\phi] = 0$ to avoid these complications.

\subsection{Background field gauge}

We close this section of the appendix with some comments on the background field gauge used for the non-Abelian gauge connection in the main text. The initial assumption that our action may be decomposed as \eqref{AFS1} is violated in this gauge as the bulk action \eqref{YM3} depends on the background $\overline A$ and the fluctuations $a$ separately, not only in the combinations $A$. Seemingly worse, a variation with respect to $\overline A$ would appear to result in bulk operator insertions which are not simply proportional to the equations of motion.

However, these issues are easily avoided by using the invariance of physical quantities under changes of the gauge-fixing function. In the bulk Minkowski spacetime, the Carrollian partition function is computing the transition amplitude between two on-shell gluon coherent states. So long as we only consider external states where the gluon indeed becomes asymptotically free, i.e. when gluon scattering makes sense to begin with, on-shell multi-gluon states are physical and thence the Carrollian partition function is independent of the gauge-fixing function. We give some further comments in the next subsection, appendix \ref{APP: gauge invariance}

So we could replace the background field gauge by, say, the Lorenz gauge-fixing function $f^a = \p^\mu A_\mu^a$ for the purposes of working with an action that can be decomposed as in \eqref{AFS1}. In this gauge, the above argument proceeds without issue, and one can revert to background field gauge knowing that the Carrollian partition function computes the same quantities computed by LSZ reduction.

\subsection{Gauge invariance}
\label{APP: gauge invariance}

A primary motivation behind the original S-matrix path integral proposal \cite{Arefeva:1974jv} was in the handling of gauge redundancy, since  computations could be carried out in a manifestly on-shell manner. We found it illuminating to review how some approaches to gauge redundancy connect to the S-matrix path integral formalism.

One should first note that the question at hand is the gauge invariance of amplitudes with external gluons, i.e. the independence of these amplitudes from the choice of gauge-fixing function. The argument that correlators of gauge invariant operators are independent of the gauge-fixing function is simple and can be found in textbook treatments, e.g. \cite{Weinberg:1996kr}. However, the standard procedure to compute an amplitude with external gluons first computes the bulk correlator of gauge field insertions. This object is generally dependent on the gauge-fixing function, and one must argue that any such dependence drops out when applying the LSZ reduction.

The S-matrix path integral avoids this intermediate gauge-dependent object, but upon further thought does not dodge the basic question: are asymptotic gluon states elements of the physical, and hence gauge invariant, Hilbert space?

One can approach this question by constructing direct diagrammatic arguments, for example \cite{Abbott:1983zw} does so for the particular case of background field gauge. Alternatively, one can argue that in the interaction picture the Gauss law constraint linearizes, so on-shell coherent gluon states, and hence on-shell multi-particle gluon states, lie in the physical Hilbert space \cite{Faddeev:1980be}.\footnote{The same reference also contains a diagrammatic argument in the covariant gauges.}

Approaching gauge invariance by way of BRST symmetry provides a window into the field theoretic assumptions required for gluon amplitudes to be independent of the gauge-fixing function.\footnote{See \cite{Kugo:1979gm} for similar comments, but from the perspective of LSZ reduction.} If we integrate in the auxiliary Gaussian field $h^a$, the BRST transformations may be written
\eq{GI1}{
    \delta_B A_\mu^a = D_\mu c^a,\ \ \ \delta_B c^a = -\frac{1}{2}\struct{a}{b}{c}c^bc^c,\ \ \ \delta_B \overline c^a = h^a,\ \ \ \delta_B h^a = 0.
}
Note that we define the BRST transformation with respect to the gauge field $A_\mu^a$ before shifting to $A_\mu^a = \overline A_\mu^a + a_\mu^a$. The BRST operator is nilpotent, $\delta_B^2 = 0$, and the bulk action \eqref{YM3} may be written explicitly in the form
\eq{GI2}{
    I_\text{bulk} = -\frac{1}{4}\int\dr^4x F^a_{\mu\nu}F_a^{\mu\nu} + \delta_B\int\dr^4 x \Psi
}
where $\Psi$ contains all the dependence on the choice of gauge-fixing function, ghosts, and auxiliary field. Hence the bulk action is manifestly BRST invariant, meaning the time evolution operator is BRST invariant. It follows that if the boundary terms \eqref{YM7} are also BRST invariant, the coherent gluon states prepared on $\scrI^\pm$ by our choice of Carrollian data must also be BRST invariant, and hence elements of the physical Hilbert space. As such, the Carrollian partition function as a whole will be independent of our choice of gauge-fixing function.

Since the transformation of $A_\mu^a$ is non-linear, the BRST transformation does not generically respect the split into positive and negative frequencies. In particular, the unfixed frequency content of $c^a$ may freely vary near $\scrI^\pm$ and so there will be regions in the ghost integration domain in which the boundary terms \eqref{YM7} are not BRST invariant. However, this is only a potential issue non-perturbatively. So long as we work perturbatively around the free theory, the $c^a$ integral is dominated by fluctuations around the solutions to the free massless scalar wave equation. All such fluctuations go as $1/r$ near null infinity, and hence $\delta_B A_A^a = D_A c^a$ starts at order $1/r$, so the leading component of the field, $A_{0A}^a$, which makes an appearance in the boundary term \eqref{YM7} is left invariant by BRST transformations.

Of course, we also require our external states to have zero ghost number, so it is necessary that a BRST transformation cannot turn on non-trivial boundary data for $c^a$ or $\overline c^a$. We see from \eqref{GI1} that $\delta_B c^a \sim 1/r^2$ is subleading to the boundary data of $c^a$. But for the antighost we need to impose on $\scrI^+$ that $\delta_B \overline c_a^- = h_a^-$ fall off faster than $1/r$, and similarly for $c_a^+$ on $\scrI^-$. But upon integrating out the auxiliary field, $h_a^- = (\overline D_\mu A^\mu_a)^- = \p_\mu A^\mu_{-a} + g\struct{a}{b}{c}(\overline A_\mu^b A^\mu_c)^-$. Since we choose the background $\overline A_\mu^a$ to behave as a free field, near null infinity its Cartesian components behave as $1/r$.

Using again that we work in perturbation theory so $A_\mu^{a}$ is dominated by the behavior of fluctuations about the free massless wave equation, it also goes as $1/r$, and hence to leading order $h_a^- \approx \p^\mu A_\mu^{a-} \sim 1/r$. It follows that BRST invariance of the zero ghost external state requires we impose $\p^\mu A_\mu^{a-} = 0$ at leading order near null infinity. This condition is obeyed by the bulk-boundary propagator \eqref{YM20}, and hence the perturbatively computed correlators of the conjugate operators also satisfy this condition.

Finally, we note that this argument suggests an interpretation of large gauge transformations within the BRST framework: they are the class of gauge parameters whose behavior cannot be captured by the dynamical ghost $c^a$. In the present case, since the ghosts behaving as $1/r$ dominate, gauge transformations behaving as $r^0$ can act on components of the field which are physical, i.e. left invariant by BRST transformations.

\section{Toy model 3-point function}
\label{Sec:Toy model 3-point function}

Here we perform the integrals in \eqref{SSS5} explicitly. We define $y = (u_y, \hat y)$, $y_1 = (u_1, \hat y_1)$, and $y_2 = (v_2, \hat y_2)$ throughout. We begin with the integral on the first line which originates from the first diagram in \eqref{SSS2},
\eq{3pt1}{
    I_F &= \int_{M^4}K_{+\mu}^A(x; y)\big(\p_x^\mu G_F(x; x_0) K_+(x; y_1) - G_F(x; x_0) \p_x^\mu K_+(x; y_1) \big)\cr
    &= -2\int_{M^4}K^A_{+\mu}(x; y) \p_x^\mu K_+(x; y_1)G_F(x; x_0)
}
where in the second line we have used that $K^A_{+\mu}$ defined in  \eqref{YM20} obeys $\p^\mu K^A_{+\mu}=0$, and that the boundary term arising from the integration by parts vanishes. To evaluate this integral it will be useful to use the Fourier representations of the propagators,
\eq{3pt2}{
    G_F(x; x_0) &= -i\int \frac{\dr^4 p}{(2\pi)^4}\frac{e^{ip\cdot(x - x_0)}}{p^2 - i\epsilon},\cr
    K_\pm(x; x') &= -\frac{i}{(2\pi)^2}\int_0^\infty \dr\omega \omega e^{\mp i\omega(u' + n(\pm \hat x')\cdot x \mp i\epsilon)}.
}

With this,
\eq{3pt3}{
    I_F &= -2\frac{n(\hat y_1)\cdot \epsilon^\alpha(\hat y)\hat\epsilon_\alpha^{*A}(\hat y)}{(2\pi)^4}\int\dr^4 x \frac{\dr^4 p}{(2\pi)^4}\omega_1\dr\omega_1 \omega_2^2\dr\omega_2 e^{-i\omega_1(u_y - i\epsilon)} e^{-i\omega_2(u_1 - i\epsilon)}\frac{e^{-ip\cdot x_0}}{p^2 - i\epsilon}e^{i(p - \omega_1 n(\hat y) - \omega_2 n(\hat y_1))\cdot x}\cr
    &= -2\frac{n(\hat y_1)\cdot \epsilon^\alpha(\hat y)\hat\epsilon_\alpha^{*A}(\hat y)}{(2\pi)^4}\int\omega_1 \dr\omega_1 \omega_2^2\dr\omega_2 \frac{e^{-i\omega_1(u_y + n(\hat y)\cdot x_0 - i\epsilon)} e^{-i\omega_2(u_1 + n(\hat y_1)\cdot x_0 - i\epsilon}}{(\omega_1 n(\hat y) + \omega_2 n(\hat y_1))^2 - i\epsilon}.
}
But now since $(\omega_1 n(\hat y) + \omega_2 n(\hat y_1))^2 = 2\omega_1 \omega_2 n(\hat y)\cdot n(\hat y_1)$, we can use the freedom to redefine $\epsilon$ together with $\omega_1,\omega_2> 0$ to perform the remaining elementary integrals and find the result,
\eq{3pt4}{
    I_F = -\frac{1}{(2\pi)^2}\frac{n(\hat y_1)\cdot \epsilon^\alpha(\hat y)\hat\epsilon^{*A}_\alpha(\hat y)}{n(\hat y)\cdot n(\hat y_1) - i\epsilon} \frac{1}{u_y + n(\hat y)\cdot x_0 - i\epsilon}K_+(x_0; y_1)
}
where we have identified the scalar bulk-boundary propagator in the result.

The computation for
\eq{3pt5}{
    I_P &= \int_{M^4}\dr^4 x K_{+\mu}^A(x; y)\big( \p_x^\mu K_-(x; y_2) G_F(x; x_0) - K_-(x; y_2) \p_x^\mu G_F(x; x_0) \big)\cr
    &= 2\int_{M^4}\dr^4 x K_{+\mu}^A(x; y)G_F(x; x_0)\p_x^\mu K_{-}(x; y_2)
}
is essentially the same, but with some sign flips. We write
\eq{3pt6}{
    I_P &= -2\frac{n(-\hat y_2)\cdot \epsilon^\alpha(\hat y)\hat\epsilon_\alpha^{*A}(\hat y)}{(2\pi)^4} \int\dr^4 x \frac{\dr^4 p}{(2\pi)^4}\omega_1 \dr\omega_1 \omega_2^2\dr\omega_2 e^{-i\omega_1(u_y - i\epsilon)} e^{i\omega_2(v_2 + i\epsilon)} \frac{e^{-ip\cdot x_0}}{p^2 - i\epsilon} e^{i(p - \omega_1 n(\hat y) + \omega_2n(-\hat y_2))\cdot x}\cr
    &= -2\frac{n(-\hat y_2)\cdot \epsilon^\alpha(\hat y)\hat\epsilon_\alpha^{*A}(\hat y)}{(2\pi)^4} \int\omega_1 \dr\omega_1 \omega_2^2\dr\omega_2 \frac{e^{-i\omega_1(u_y + n(\hat y)\cdot x_0 - i\epsilon)} e^{i\omega_2(v_2 + n(-\hat y_2)\cdot x_0 + i\epsilon)}}{(\omega_1 n(\hat y) - \omega_2 n(-\hat y_2))^2 - i\epsilon}.
}
As before, we expand $(\omega_1 n(\hat y) - \omega_2n(-\hat y_2))^2 = -2\omega_1 \omega_2 n(\hat y) \cdot n(-\hat y_2)$ and perform the remaining integrals to find
\eq{3pt7}{
    I_P = \frac{1}{(2\pi)^2}\frac{n(-\hat y_2)\cdot \epsilon^\alpha(\hat y) \hat\epsilon^{*A}_\alpha(\hat y)}{n(\hat y)\cdot n(-\hat y_2) + i\epsilon} \frac{1}{u_y + n(\hat y)\cdot x_0 - i\epsilon}K_-(x_0; y_2).
}

\section{Conditionally convergent integrals}
\label{Sec: Conditionally convergent integral}

Our derivation of the soft theorem involves a subtle factor of two having to do with the evaluation of certain conditionally convergent integrals.  In this appendix we establish a general result involving such integrals.

Let $f(z)$ be a function analytic in the lower half-plane including the real axis, with leading behavior $f(z) \sim {1\over z}$ as $z\rt \infty$.  We also assume that there exists an $R$ such that  $f(z)$ is analytic for $|z|>R$.   We define  the following two integrals
\eq{kl1}{  I_1& = \lim_{L\rt \infty}\int_{-L}^L f(z) dz  \cr
  I_2& = \lim_{\omega \rt 0^+}   \lim_{L\rt \infty}\int_{-L}^L f(z)  e^{i\omega z} dz~, }
  where $\omega$ is real and positive.   We will prove that 
  \eq{kl2}{ I_2 = 2I_1~.}
A simple example is $f(z) = {1\over z-i\eps}$, for which the validity of \rf{kl2} is easily established by elementary evaluation.  

For the general case we first of all use that
  \eq{d7fb}{  \lim_{L\rt \infty}\int_{-L}^L f(z)  e^{i\omega z} dz = \lim_{L\rt \infty} \oint_{C_L}  f(z)  e^{i\omega z} dz  }
  where  $C_L$ is a closed contour consisting of the union of the segment $z\in [-L,L]$ and the semi-circular arc in the upper half plane connecting the endpoints,  with counter-clockwise orientation. \rf{d7fb} is valid since the arc doesn't contribute by  Jordan's Lemma.\footnote{Strictly speaking we need that for each semi-circular contour of radius $L$ larger than some $L_0$,  there  exists a number $M_L$ such that $|f(z)|<M_L$ for all points on the contour, and $\lim_{L\rt \infty} M_L=0$. } Next, if $L>R$, then $\oint_{C_L} f(u)  e^{i\omega u} du$ is independent of $L$ by Cauchy's theorem.    So we can  write
  \eq{d7fc}{    \lim_{L\rt \infty}\int_{-L}^L f(u)  e^{i\omega u} du = \oint_{C}  f(u)  e^{i\omega u} du  }
where $C$ is taken to be any $C_L$ with $L>R$.   Since $C$ is a bounded contour, it's clear that
\eq{d7fd}{ \lim_{\omega \rt 0^+}   \oint_{C}  f(u)  e^{i\omega u} du = \oint_{C}  f(u)  du   }
Now that the $e^{i\omega z}$ factor is absent  we can go back to thinking of $C$ as being the union of $z \in [-L,L]$ and the semi-circular arc of radius $L>R$ oriented counter-clockwise,  which we'll call $S^+_L$.   So at this point we have that
\eq{d7fd2}{ I_2 = I_1 + \lim_{L\rt \infty} \int_{S^+_L}  f(z)  dz }
  Now, since $f(z)$ is analytic in the lower half-plane we know from Cauchy's theorem that.
  \eq{d7fe}{ I_1 =  \lim_{L\rt \infty} \int_{S^-_L}  f(z)  dz   }
where $S^-_L$ is the semi-circular arc in the lower half-plane with counterclockwise orientation. However, we also have that
\eq{d7ff}{   \lim_{L\rt \infty} \int_{S^-_L}  f(z)  dz = \lim_{L\rt \infty} \int_{S^+_L}  f(z)  dz  }
To see this, change variables as $z=1/y$.  Then the two contours are tiny semi-circles surrounding the origin and the integrands behave as $ {c \over y}dy$ from our assumptions above.  It's clear that in the limit that the contours approach the origin that only the leading simple pole term contributes, and then  \rf{d7ff} follows.   We have thus deduced that
\eq{d7fg}{ I_2 = I_1 +I_1 =2I_1}
as desired.

As described in the main text, the Ward identity associated with large gauge transformations gives a result for an integral of the form $I_1$, while the soft limit of the amplitude involves and integral of the form $I_2$.  Obtaining the correct coefficient in the soft theorem thus requires the result \rf{kl2}.

\section{Charge algebra at finite Goldstone}
\label{Sec: Charge algebra at finite Goldstone}

Here we consider the effects of working at finite Goldstone on the charges and their algebra. In section \ref{Sec:Charge algebra} we gave the result \eqref{CA1} computed from the charges \eqref{SG3}, which are the correct charges at lowest order in the Goldstone $\Phi$. However, for some applications one might require the charges written at finite values of the Goldstone, and in principle one might also worry whether the charge algebra is deformed at finite Goldstone. In this section we address both these issues.

To begin, it will be most useful to work only on $\scrI^+$ which, in light of \eqref{ST2}, means considering variations which are not antipodally matched so that we can assign independent variations of the Goldstone to the charge on $\scrI^+$ and $\scrI^-$. Of course, one should only expect that these define a symmetry when the Goldstone is matched, but there is no issue considering the charges without matching. To do this, we simply consider $C_A$ on $\scrI^\pm$ to be independent, and we can recover the result of the antipodal matching condition by writing
\eq{FG1}{
    \frac{\delta}{\delta \Phi^a(\hat x)} = \frac{\delta}{\delta \Phi^a_+(\hat x)} + \frac{\delta}{\delta \Phi_-^a(-\hat x)}
}
where $\Phi_\pm^a$ are the Goldstones on $\scrI^\pm$.

Under a gauge transform $U = e^{ig\lambda}$, the field $C_A = -\frac{i}{g}V^{-1}\p_A V$, where $V = e^{ig\Phi}$, transforms as
\eq{FG1a}{
    igC_A' = U^{-1} V^{-1}\p_A V U + U^{-1}\p_A U = (VU)^{-1}\p_A(VU),
}
so the Goldstone group element $V$ transforms under right multiplication, and hence the transformed Goldstone is determined by
\eq{FG2}{
    e^{ig\Phi'} = e^{ig\Phi}e^{ig\lambda}.
}
The variation in $\Phi$ under a gauge transformation can then be determined order by order in $\Phi$ by expanding \eqref{FG2} to lowest order in $\lambda$,
\eq{FG3}{
    \delta_\lambda \Phi &= \lambda + \frac{1}{2}[ig\Phi, \lambda] + \frac{1}{12}[ig\Phi, [ig\Phi, \lambda]] + \cdots.
}

Working now only on $\scrI^+$, we may obtain the finite Goldstone analogs of \eqref{YM9} and \eqref{YM11} by considering the variation \eqref{FG3} applied only to Goldstone. On the one hand,
\eq{FG4}{
    \delta_\lambda\ln Z = \int\sqrt{\gamma}\dr^2x \delta_\lambda\Phi^a_+(\hat x)\frac{\delta}{\delta \Phi^a(\hat x)}\ln Z.
}
On the other hand, using the variation \eqref{YM6} we have
\eq{FG5}{
    \delta_\lambda\ln Z &= i\int_{\scrI^+}\sqrt{\gamma}\dr^3 x \tr\p_u a^+_{0A}(u, \hat x)\delta_\lambda C^A(\hat x)\cr
    &= i\int_{\scrI^+}\sqrt{\gamma}\dr^3 x \tr\p_u a^+_{0A}(u, \hat x)D^A[C]\lambda\cr
    &= -\frac{1}{2}\int_{\scrI^+}\sqrt{\gamma}\dr^3 x\tr D_A[C]\frac{\delta}{\delta \hatline{A}\subsuper{{0A}}{-}(u, \hat x)}.
}
Here we have defined the covariant derivative with respect to the Goldstone's flat connection on the sphere, $C_A$, by $D_A[C]$ and used that the gauge variation of a connection is the covariant derivative of the gauge parameter. Furthermore, note that because we defined the hard/soft split as in \eqref{YM2}, the hard variations are unchanged by the presence of a finite Goldstone. Together these give the finite Goldstone relation between soft and hard variations,
\eq{FG6}{
    \int_{\scrI^+}\sqrt{\gamma}\dr^2 x \delta_\lambda\Phi^a(\hat x)\frac{\delta}{\delta\Phi^a(\hat x)} = -\frac{1}{2}\int_{\scrI^+}\sqrt{\gamma}\dr^3 x \lambda^a(\hat x) D^{ab}_A[C]\frac{\delta}{\delta \hatline{A}\subsuper{0A}{b-}(u, \hat x)}.
}

It is straightforward to see from \eqref{ST1} that the charge $\hat Q_+[\lambda]$ is given by
\eq{FG7}{
    i\hat Q_+[\lambda] = \int_{\scrI^+}\sqrt{\gamma}\dr^2 x\delta_\lambda \Phi^a(\hat x)\frac{\delta}{\delta \Phi^a(\hat x)} + g\struct{a}{b}{c}\int_{\scrI^+}\sqrt{\gamma}\dr^3 x\lambda^a(\hat x)\hatline{A}\subsuper{0A}{b-}(u, \hat x)\frac{\delta}{\delta\hatline{A}\subsuper{0A}{c-}(u, \hat x)}.
}
In this form, it's clear that $i\hat Q_+[\lambda]$ is nothing but the phase space vector field generating the action of a large gauge transform, and hence
\eq{FG8}{
    \relax[\hat Q_+[\lambda], \hat Q_+[\rho]] = i\hat Q_+[[\lambda, \rho]].
}
Note that we would not have been able to find this result if considered only the lowest order contributions to the charge in the Goldstone since in those terms there would be nothing for the Goldstone variations to act upon.

The calculation performed here is somewhat different than the one considered in the main text, where the analog of \eqref{FG6} was used to replace all soft variations with hard ones. At finite Goldstone, this would mean asking whether the charges
\eq{FG9}{
    i\hat Q'_+[\lambda] = -\frac{1}{2}\int_{\scrI^+}\sqrt{\gamma}\dr^3x\lambda^a D^{ab}_A[C] \frac{\delta}{\delta\hatline{A}\subsuper{0A}{b-}} + g\struct{a}{b}{c}\int_{\scrI^+}\sqrt{\gamma}\dr^3 x\lambda^a(\hat x)\hatline{A}\subsuper{0A}{b-}(u, \hat x)\frac{\delta}{\delta\hatline{A}\subsuper{0A}{c-}(u, \hat x)}
}
also obey the algebra \eqref{FG8}. It is simple to show that this is true at zeroth order in the Goldstone, but at finite Goldstone this calculation is harder.

We can indirectly show the result as follows. We can phrase \eqref{FG6} as the statement that the vector field
\eq{FG10}{
    i\hat R[\lambda] = \int_{\scrI^+}\sqrt{\gamma}\dr^2x\delta_\lambda\Phi^a\frac{\delta}{\delta \Phi^a} + \frac{1}{2}\int_{\scrI^+}\sqrt{\gamma}\dr^3x\lambda^a D^{ab}_A[C]\frac{\delta}{\delta\hatline{A}\subsuper{0A}{b-}}
}
annihilates the partition function, $\hat R[\lambda]Z = 0$. Hence in the space of data $(\hatline{A}\subsuper{0A}{a-},\Phi^a)$ (considering only the data on $\scrI^+$ for the moment), $Z$ is not an arbitrary functional, but rather has level curves -- submanifolds of constant $Z$ -- to which $\hat R$ is tangent.

In this language, which is only possible because $\hat Q$ and $\hat R$ happen to be vector field operators, the operators \eqref{FG9} will obey the algebra \eqref{FG8} if and only if $[\hat R[\lambda], \hat Q[\rho]] = 0$ on the subspace of functions obeying $\hat R = 0$. That is, if the commutator depends only on $\hat R$.\footnote{To see this, note $\hat Q'_+[\lambda] = \hat Q_+[\lambda] - \hat R[\lambda]$ and so $[\hat Q_+'[\lambda], \hat Q_+'[\rho]] = [\hat Q_+[\lambda], \hat Q_+[\rho]] - [\hat Q_+[\lambda], \hat R[\rho]] - [\hat R[\lambda], \hat Q_+[\rho]] + [\hat R[\lambda], \hat R[\rho]]$. The commutator of $\hat R$'s automatically annihilates the any $Z$ obeying $\hat R Z = 0$. Strictly, one only requires $[\hat R[\lambda], \hat Q_+[\rho]] - (\lambda\leftrightarrow\rho)$ to annihilate the same class of functions $Z$, but we will see that $[\hat R[\lambda], \hat Q_+[\rho]]$ is already antisymmetric under $\lambda\leftrightarrow\rho$.}

We can use the fact that $\hat Q_+[\rho]$ is the generator of large gauge transforms to note that $[\hat R[\lambda], \hat Q[\rho]]$ is nothing but the gauge transformation of the dynamical parts of $\hat R[\lambda]$. But if the gauge parameter $\lambda$ were transformed simultaneously, $\hat R[\lambda]$ would be gauge invariant since there are no free indices. Hence we see
\eq{FG11}{
    \relax[\hat R[\lambda], \hat Q_+[\rho]] = i\hat R[[\lambda, \rho]].
}
Note that the same argument could have been made to obtain \eqref{FG8} without relying upon already knowing the algebra of the vector fields.

We now see that on the space of functions obeying $\hat R[\lambda] Z = 0$, the commutator \eqref{FG11} indeed vanishes, and so the charges \eqref{FG9} indeed obey the same algebra \eqref{FG8}, on the space of functions obeying \eqref{FG6}, for finite Goldstone, extending the direct computation which showed \eqref{CA1}.

We note that these arguments also make clear that the charges \eqref{FG9} will only act in the same way as the charges \eqref{FG7} if one properly solves the constraint \eqref{FG6}. That is, if one uses \eqref{FG6} to solve for the Goldstone as a functional of the hard data. More practically, to use \eqref{FG9} to act on a functional of the Goldstone, the hard variation of a Goldstone should be defined by \eqref{FG6}.

\section{Massive particles}
\label{Sec:Massive particles}

In this appendix we make some preliminary comments about including massive particles in our story.\footnote{For previous relevant work see \cite{Kapec:2015ena,Campiglia:2015qka,Have:2024dff}.}  The asymptotic data for a theory of massive particles lies at past and future timelike infinity.  Focusing on future infinity $i^+$, we employ a hyperbolic slicing in order to ``resolve" $i^+$  \cite{Beig:1982ifu,Winicour:1988aq,Ashtekar:1991vb,deBoer:2003vf}.   If $(t,r)$ denote the coordinates appearing in $ds^2 = -dt^2+ dr^2+ r^2 d\Omega^2$ we define 
\eq{mm1}{\tau^2 = t^2-r^2~,\quad \rho = {r\over \sqrt{t^2-r^2}} }
so that 
\eq{mm2}{ ds^2 &= -d\tau^2+ +\tau^2 \left[ {d\rho^2 \over 1+\rho^2}+\rho^2 d\Omega^2\right] \cr
& =  -d\tau^2 +\tau^2dH_3^2 }
The future hyperboloid $H_3$ is the natural home for asymptotic data.  Lorentz transformations clearly map an $H_3$ at fixed $\tau$ to itself. If we introduce stereographic coordinates $(z,\zb)$ on the two-sphere, then SL(2,$\mathbb{C}$) Lorentz transformations act as 
\eq{mm3}{ z\rt {az+b\over cz+d}~,\quad  \zb\rt {\bar{a} z+\bar{b}\over \bar{c} z+\bar{d} }  }
as $\rho \rt \infty$ on a fixed $H_3$. On the other hand, translations act as a $(\rho,z,\zb)$ dependent shift of $\tau$.  

The Lorentz invariant hyperboloid $H_3$ may be identified with the mass-shell hyperboloid $\omega^2 = \vec{p}^{\,\,2}$, providing a natural identification between on-shell momenta and points on $H_3$.  For instance, the integral appearing in the mode expansion of the free field operator 
\eq{mm4}{ \phi(x) = \int\! {d^3p \over (2\pi)^3}{1\over 2\omega_p}\big( b(\pv) e^{ipx} + b^\dagger(\pv)e^{-ipx}\big)~,\quad \omega_p =p^0 = \sqrt{\pv^2+m^2}  }
may be reinterpreted as an integral over $H_3$ with coordinates $p_i$, and with boundary data specified by the negative frequency modes $b^\dagger(\vec{p})$.  To elaborate, a large $\tau$ saddle point analysis gives the asymptotic behavior of the negative frequency part of the field as 
\eq{mm5}{ \phi_-(\tau,y) &\approx {e^{ im\tau }\over \tau^{3/2} } \overline\phi{_-^{\,3/2}}(y)}
with
\eq{mm6}{
\overline\phi{_-^{\,3/2}} (y)& =   {\sqrt{m} \over 2(2\pi) ^{3/2}} e^{3i\pi /4}   b^\dagger (m\rho\yh)    }
Here $y=(\rho,\yh)$ denotes a point on $H_3$, with the identification with massive particle momenta given by $\vec{p} = m \rho \hat{y}$.  We can use this to extract the bulk-boundary propagator.  In particular we have 
\eq{mm7}{ \phi(x) = \int_{H_3} d^3y \sqrt{g_{H_3}} K_{i^+}(x,y)  \overline\phi{_-^{\,3/2}}(y)  }
with 
\eq{mm8}{ K_{i^+}(x,y) =  {m^{3/2} \over (2\pi)^{3/2}} e^{-3i\pi/4}e^{i   m\tau \big( \sqrt{1+\rho'^2}\sqrt{1+\rho^2}-\rho' \rho \xh \cdot \yh \big)   }~.   } 
The bulk-boundary propagator may also be related to the boundary limit of massive scalar Feynman propagator as
\eq{mm9}{ K_{i^+} (x_2,y_1) \approx  2m\tau_1^{3/2} e^{im \tau_1} G_F(x_2,x_1)~,\quad \tau_1 \rt \infty }
Analogous formulas apply to past timelike infinity.

Using these expressions one can compute a boundary partition function by evaluating flat space Witten diagrams, in direct analogy with the massless case.  While all of this is quite straightforward, what is less clear is how to most usefully combine the massive and massless theories into a  unified description, for example in terms of a theory that lives --- in some sense --- on the union of timelike and null infinity.  This is left for future work.

\bibliographystyle{bibstyle2017}
\bibliography{collection}

\end{document}